\documentclass[aps,pra,reprint, amsmath, amssymb,floatfix,superscriptaddress]{revtex4-1}
 
\usepackage{bm}
\usepackage{bbm}
\usepackage[retainorgcmds]{IEEEtrantools}
\usepackage{graphicx}
\usepackage{mathrsfs}
\usepackage{amsmath}
\usepackage{amsfonts}
\usepackage{amssymb}
\usepackage{color,xcolor}
\usepackage{times,txfonts}
\usepackage{nicefrac}
\usepackage{ragged2e}
\usepackage{float}
\usepackage{braket}
\usepackage{tikz}

\usepackage[colorlinks=true,linkcolor=blue,urlcolor=blue,citecolor=blue,pdfusetitle]{hyperref}

\usepackage{blindtext}

\newcommand{\geff}{g_{\rm eff}}

\begin{document}

\title{Stochastic trajectories and excursions in a double quantum dot system}
\date{\today}
\author{Guilherme Fiusa}
\email{gfiusa@ur.rochester.edu}
\affiliation{Department of Physics and Astronomy, University of Rochester, Rochester, New York 14627, USA}
\affiliation{Center for Coherence and Quantum Science, University of Rochester, Rochester, New York 14627, USA}
\author{Pedro E. Harunari}
\affiliation{Aix Marseille Université, CNRS, CINAM, Turing Center for Living Systems, 13288 Marseille, France}
\author{Alberto J. B. Rosal}
\affiliation{Department of Physics and Astronomy, University of Rochester, Rochester, New York 14627, USA}
\affiliation{Center for Coherence and Quantum Science, University of Rochester, Rochester, New York 14627, USA}
\author{John M. Nichol}
\affiliation{Department of Physics and Astronomy, University of Rochester, Rochester, New York 14627, USA}
\affiliation{Center for Coherence and Quantum Science, University of Rochester, Rochester, New York 14627, USA}
\author{Gabriel T. Landi}
\affiliation{Department of Physics and Astronomy, University of Rochester, Rochester, New York 14627, USA}
\affiliation{Center for Coherence and Quantum Science, University of Rochester, Rochester, New York 14627, USA}

 \begin{abstract}
We investigate the trajectory-level stochastic dynamics of a double quantum dot
system using the formalism of stochastic excursions. An excursion is a
sub-trajectory that begins when the system leaves a chosen reference state (or more generally, a set of states)---here,
the empty double dot---and ends upon its first return to it. Filtering trajectories
into excursions extends full counting statistics and gives access to the
correlations between thermodynamic currents and excursion durations. The main
objects of study are counting observables, defined as weighted sums of transition
counts within a single excursion. For three main counting observables---charge
current, dynamical activity, and entropy production---we compute averages and noise
contributions and show how they provide insights into the operation of the double
quantum dot system. At the trajectory level, we analyze outcome distributions for
transport and connect the results with trade-offs between successful and
unsuccessful events that shape the overall performance of the double quantum dot as
a transport device. We then discuss the populations of the two dots, as well as
their correlations. Finally, we illustrate how the framework provides insights into
the thermodynamics of precision through the Fano factor and thermo-kinetic
uncertainty relations, where different regimes are fundamentally constrained either
by entropy production or by dynamical activity. Altogether, our work is a case study
that highlights the utility of the excursion framework as a toolkit to analyze many
quantities of interest and to uncover the structure of nonequilibrium fluctuations.
Moreover, it also suggests new avenues for refining uncertainty relations and
understanding transport in mesoscopic systems.
 \end{abstract}

\maketitle{}

\section{Introduction}

In many experiments, especially involving nanoscale and quantum processes, one only has limited access to an experiment, through noisy and imperfect data. 
In the context of transport and thermodynamics, the data is often some kind of thermodynamic current, such as charge, heat or work. 
Investigations often focus on steady-states,
however, that only provides access to average currents. 
To study current fluctuations, one must go beyond steady-states and analyze the problem at the level of individual stochastic trajectories, where each realization of the dynamics composes a stochastic process, which can then be analyzed using the toolbox of stochastic thermodynamics~\cite{Seifert2012, Strasberg2017, seifertEntropyProductionStochastic2005a, Esposito2009, Campisi2011, Campbell2025,landi2024a, Saito2008, Alhassid2000}.

Full access to a stochastic trajectory, however, is often experimentally prohibitive. 
This has motivated various approaches aimed at capturing different aspects of the problem, such as coarse graining~\cite{espositoStochasticThermodynamicsCoarse2012, rahavFluctuationRelationsCoarsegraining2007,Tan2021, Gomez_Marin2008, Puglisi2010, Teza2020, Bo2014} and statistical inference~\cite{harunariWhatLearnFew2022, Maier2025, 
JannaPRR2024, boEntropyProductionStochastic2014, biskerHierarchicalBoundsEntropy2017, yuDissipationLimitedResolutions2024a, espositoStochasticThermodynamicsCoarse2012, ehrichTightestBoundHidden2021, blomMilestoningEstimatorsDissipation2024, liangThermodynamicBoundsSymmetry2024}.
Although  successful, those approaches shed little insight into how noise is actually generated at the trajectory level. 
Motivated by this, some of us have recently proposed a framework for describing the full counting statistics of individual stochastic excursions~\cite{Fiusa2025, Fiusa2025a}.
An excursion is defined as a specific filtering of a stochastic trajectory, where the state space is broken into two regions, labeled $A$ and $B$. An excursion is a sub-trajectory that begins with a transition $A \to B$ and ends with another $B \to A$. 
The key usefulness of this formalism is the ability to study the statistics of counting observables and thermodynamic currents at the level of individual excursions. 
Moreover, the formalism allows the representation of steady-state average and fluctuations in terms of quantities pertaining to counting observables within individual excursions, as well as quantities pertaining to excursion times.

The goal of the present paper is to showcase the usefulness of this formalism by applying it to an experimentally relevant model; namely, electronic transport across a double quantum dot (DQD) system. Transport in multiple quantum dots~\cite{van2002electron} has many potential applications at the intersection of quantum and classical physics, including charge pumping~\cite{Croy2012, Mart_nez_Mares2004,mills2019shuttling}, microwave photon detection~\cite{oppliger2026tunable,khan2021efficient}, quantum simulation~\cite{hensgens2017quantum}, qubit initialization~\cite{volk2019loading}, and readout ~\cite{ono2002current,johnson2005singlet}. 
While related properties have already been investigated in many contexts, ranging from statistics to random matrix theory~\cite{Alhassid2000, Beenakker1997}, 
a characterization of thermodynamic observables at the trajectory level for quantum dot transport remains open. 
In this paper, by providing a case study for the stochastic excursion framework we will, therefore, also provide a comprehensive characterization of quantum dot dynamics. 

The organization of this paper is as follows. In Sec.~\ref{sec:formalism} we review the double quantum dot model. In Sec.~\ref{sec:trajectories} we introduce the formalism of stochastic excursions and the role played by counting observables, together with the relations that connect excursion-level quantities to the more familiar steady-state current and noise. Sec.~\ref{sec:results} then provides a broad numerical characterization of the double quantum dot. The first result we discuss, in Sec.~\ref{subsec:excursion-duration}, concerns the timescales of the model. As we shall see later, timescales largely dictate the underlying dynamics. In Sec.~\ref{subsec:current-and-noise} we analyze the average current and the noise of our three main counting observables: the particle current, the dynamical activity, and the entropy production. In Sec.~\ref{subsec:pop} we discuss the steady-state populations of the two dots, as well as the correlations between them. In Sec.~\ref{subsec:suc-fail-dis} we characterize the distribution of outcomes of counting observables within individual excursions, classifying them as successes, failures, and disasters, and show how this provides new insights into the quantum dot dynamics at the trajectory level. We then turn to the thermodynamics of precision. In Sec.~\ref{subsec:tur} we illustrate one of the main results of the excursion framework, namely the decomposition of the noise in terms of excursion-level quantities, which we analyze through the Fano factor. And in Sec.~\ref{ssec:TKUR} we discuss different thermo-kinetic uncertainty relations, illustrating how the noise decomposition and the fundamental limitations over precision provide valuable insights. Conclusions are presented in Sec.~\ref{sec:conclusion}. Analytical results under the Coulomb blockade regime are collected in Appendix~\ref{app:analytical}, conservation laws and the proportionality between thermodynamic currents are established in Appendix~\ref{app:conservation}.
\

\section{The double quantum dot model}
\label{sec:formalism}

\begin{figure}
    \centering
    \includegraphics[width=1.1\linewidth]{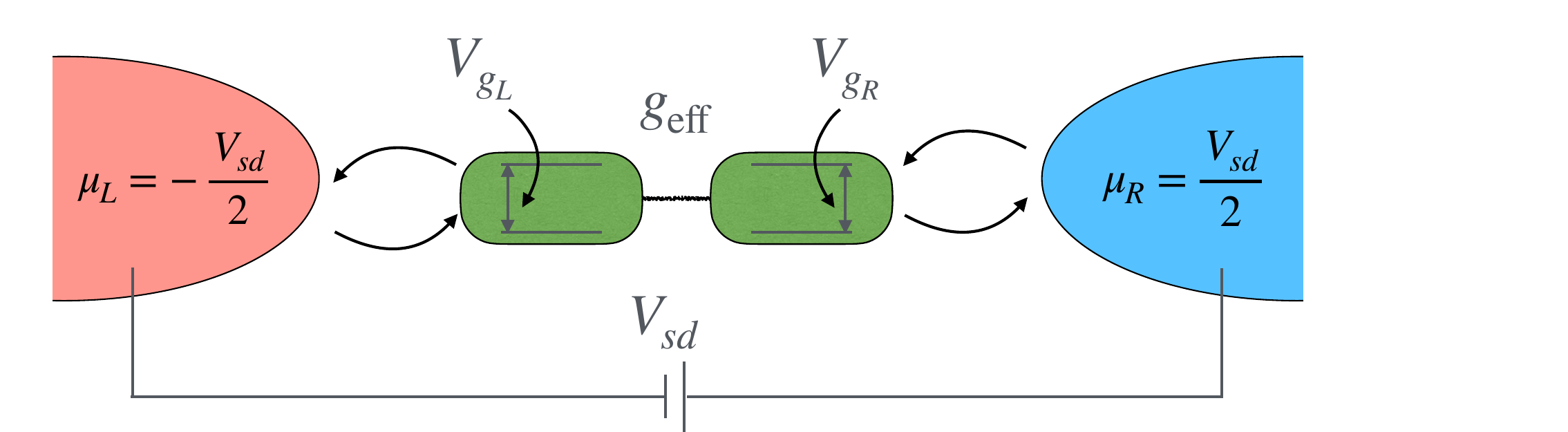}
    \caption{Schematic diagram of the double quantum dot with left and right reservoirs. 
    The energy required to add an electron to each dot depends on a gate voltage $V_{g_L}$ for the left dot and $V_{g_R}$ for the right dot.
    Two reservoirs, also labeled left and right, have a gradient of chemical potentials due to the source-drain bias voltage $V_{sd}$.
    The arrows indicate possible transitions of the system, and the dots are coupled via an effective parameter $g_{\rm eff}$.}
    \label{fig:diagram-dqd}
\end{figure}

We consider the double quantum dot (DQD) setup shown schematically in Fig.~\ref{fig:diagram-dqd}.
It consists of two quantum dots, modeled as two-level systems, that interact with an effective tunneling amplitude $g_{\rm eff}$. Each dot is connected to a different fermionic reservoir,   with chemical potentials $\mu_L$ and $\mu_R$.
The quantum dots are subject to experimentally tunable  gate voltages $V_{g_L}$  and $V_{g_R}$, which effectively set their energy gaps. 
The two dots are connected to respective fermionic leads set at the same inverse temperature $\beta_L = \beta_R = \beta$.
To induce transport, a source-drain voltage bias $V_{sd}$ is applied, inducing a gradient of chemical potentials  $\mu_L = -V_{sd}/2$ and  $\mu_R = V_{sd}/2$. 

Although the system is inherently quantum, at the high temperature limit it can be well approximated by a classical Markov master equation, a scenario that has been employed in various recent experiments~\cite{Wadhia2025, albrecht2026multilevelchargefluctuationssisige, aamirThermallyDrivenQuantum2025}.
Many DQD experiments work in the Coulomb blockade regime, whereby the electrostatic repulsion is so strong that effectively they are never occupied simultaneously.
To keep our discussion broad, in this paper we do not assume the Coulomb blockade regime, but instead consider it as a specific limit, when appropriate.

The description in terms of a classical master equation can then be constructed as follows.
We assume the system may occupy a discrete set of four states: $\ket{00}$ (both dots empty), $\ket{10}$ (left dot occupied), $\ket{01}$ (right dot occupied), and $\ket{11}$ (both dots occupied). 
Those states evolve stochastically following a Markovian classical master equation:
\begin{equation}\label{M_vec}
    \frac{d \ket{p}}{dt} = \mathbb{W}\ket{p},
\end{equation}
where $\ket{p}$ is a vector with components $p_x$ ($x=00,10,01, 11$) and 
\begin{equation}\label{mathbb_W}
    \mathbb{W} = W- \Gamma = \begin{cases}
        W_{xy} & x\neq y \\
        -\Gamma_x & x = y.
    \end{cases}
\end{equation}
denotes the stochastic transition matrix.
Here $W_{xy}$ denotes the transition rates (probability per unit time) for transitions from $y$ to $x$. 
And $\Gamma_x = \sum_{y} W_{yx}$ denote the inverse residence times the system spends in each state $x$. 
Note that we always define $W$ to be a matrix with zero in the diagonals. 
Conversely, $\mathbb{W}$ is the matrix with negative diagonals and whose columns add up to zero. 
Moreover, the ket notation for the vector state $\ket{p}$ is just a matter of convenience and it is not related to quantum mechanical energy states; i.e., $\ket{p}$ is just a shorthand for a column vector.

The system has two sources of transitions. The first are due to the fermionic reservoirs whose transition rates are given by $\gamma f_a$ for injections and $\gamma (1-f_a)$ for extractions, for $a = L, \ R$, where $f_a = (\exp{\beta} (V_{g_{a}} - \mu_a)+1)^{-1}$ denotes the Fermi function for inverse temperature $\beta$ and chemical potential $\mu_a$; $\gamma$ is the coupling strength between the dot and the reservoir.
If one of the two dots is already occupied, the transition rates have to be modified to $\Tilde{f}_a = (\exp{\beta (V_{g_{a}}+U-\mu_a)}+1)^{-1}$, due to the Coulomb repulsion energy $U$ between the electrons in each dot.
The Coulomb blockade regime corresponds to $U\to \infty$, which means both dots cannot be occupied at the same time, and it can be obtained by setting $\Tilde{f}_{a} \to 0$. 
The second source of transitions is the effective tunneling between the dots, denoted $g_{\rm eff}$, which generates the transitions $\ket{10} \leftrightarrow \ket{01}$. 
This can be calculated using perturbation theory~\cite{Prech2023}, starting from the fully quantum mechanical Hamiltonian with tunneling amplitude $g$:
\begin{equation}\label{eq:g_eff}
    g_{\rm eff} = \dfrac{2g^2\gamma}{\gamma^2+(V_{g_L}-V_{g_R})^2}.
\end{equation}
Combining all ingredients, we arrive at the following transition matrix for the DQD system (see also Fig.~\ref{fig:trajectory-transitions}(b)): 
\begin{equation}\label{W-matrix-no-coulomb-no-diag}
W =
\begin{pmatrix}
0& \gamma (1-f_L) & \gamma (1-f_R)&0\\
\gamma f_L &0 & g_{\rm eff}& \gamma (1-\tilde{f}_R)\\
\gamma f_R & g_{\rm eff} & 0 & \gamma (1-\tilde{f}_L)\\
0 & \gamma \tilde{f}_R & \gamma \tilde{f}_L & 0 
\end{pmatrix}.
\end{equation}
Since the state space is irreducible, the dynamics in Eq.~\eqref{M_vec} has a unique steady state $\ket{p^{\rm ss}}$ which is the solution of $\mathbb{W}\ket{p^{\rm ss}} =0$. 
For simplicity, we will henceforth assume both dots have the same gate voltage, i.e. $V_{g_L} = V_{g_R} = V_g$.
The tunable parameters are therefore $V_g, V_{sd}, U, g, \beta$ and $\gamma$.

\section{Trajectory level behavior and stochastic excursions}
\label{sec:trajectories}

\begin{figure}
    \centering
    \includegraphics[width=1.0\linewidth]{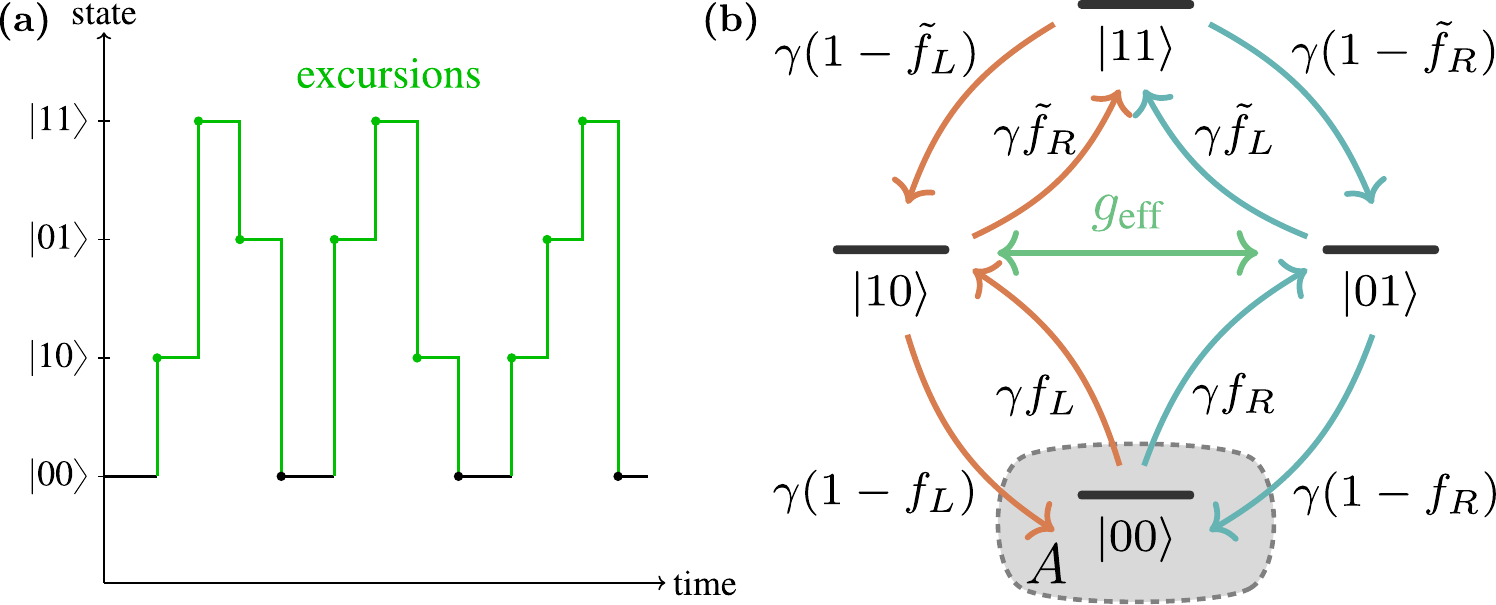}
    \caption{
    (a) Sample trajectory in the double quantum dot system, as a function of time in arbitrary units. 
    In this framework, we break the long and complicated trajectory into simpler subtrajectories dubbed excursions, indicated in green.
    (b) Energy levels of the double quantum dot model. Transition rates are displayed illustrating the physical processes. Region $A$ of the excursions is shaded grey. The transitions are illustrated with their rates.}
    \label{fig:trajectory-transitions}
\end{figure}

The master Eq.~\eqref{M_vec} describes the dynamics at the level of the ensemble. 
At the trajectory level, the system spends an exponentially distributed time in a given state before jumping on to the next one.
The stochastic trajectory is therefore described by two elements: a set of states $x_1\to x_2\to x_3\to \ldots$ that the system navigates, and a set of exponentially distributed residence times $\tau_i$ that account for how long the system spent in state $x_{i}$ before transitioning to $x_{i+1}$. 
A schematic example of said stochastic trajectory is shown in Fig.~\ref{fig:trajectory-transitions}(a). 

The steady-state of the master equation provides information on the amount of time the system spends in each state. 
More insights, however, can be obtained if we look at the statistics of trajectories themselves. 
For example, one can ask about fluctuations of thermodynamic quantities at the trajectory level~\cite{Seifert2012}.
With that in mind, here we employ the stochastic excursions formalism developed in Refs.~\cite{Fiusa2025, Fiusa2025a}, which works as follows. 
One breaks the state space $x=\{\ket{00},~ \ket{10},~ \ket{01}, ~\ket{11}\}$ in two sets, which we call $A$ and $B$ (cf. Fig.~\ref{fig:trajectory-transitions}(b)). 
An excursion is a subtrajectory that begins with a transition $A \to B$ and ends the first time the system returns from $B\to A$. 
In the process, the system jumps a random number of times on states within $B$, spending a random amount of time. 

Here we choose $A = \{\ket{00}\}$, which is a natural recurring state since it is the ground state. 
It therefore provides excursions with an intuitive interpretation. Namely, we are looking for events which start with the double dot empty; involve a certain number of tunneling events, in and out; and end when the dots become empty again. 
We therefore consider excursions to be trajectories leaving the ground state and then returning to it at a later time.
A diagram of the possible excursion paths is shown in Fig.~\ref{fig:trajectory-transitions}(b).
Both the excursion path, as well as the duration of the excursion, are random variables. 

Since $A = \{ \ket{00}\}$ the remaining region is $B = \{\ket{10}, \ket{01},\ket{11}\}$. 
Following the results in Refs.~\cite{Fiusa2025, Fiusa2025a}, we split the rate matrix $\mathbb{W}$ in Eq.~\eqref{mathbb_W} in a  block structure
\begin{equation}\label{block_W}
    \mathbb{W} = \begin{pmatrix}
\mathbb{W}_{\!\!A} & W_{\!\!AB} \\
W_{\!\!BA} & \mathbb{W}_{\!\!B}
\end{pmatrix},
\end{equation}
where $\mathbb{W}_{\!\!A}$
is $1\times 1$,
$W_{\!\!BA}$ is $3\times 1$, $W_{\!\!AB}$ is $1\times 3$ and $\mathbb{W}_{\!\!B}$ is $3\times 3$. 
In particular, since region $A$ has a single state, $\mathbb{W}_{\!\!A}= -\Gamma_A =-\gamma f_L -\gamma f_R$.
Notice that we write the off-diagonals as e.g. $W_{\!\!AB}$, instead of $\mathbb{W}_{\!\!AB}$, just to emphasize that the AB and BA blocks of $\mathbb{W}$ coincide with those of $W$ (which is not true for $\mathbb{W}_{\!\!A}$ and $\mathbb{W}_{\!\!B}$).

The matrix $W_{BA}$ contains the transitions that go from $A\to B$, and are therefore the ones that mark the start of an excursion. 
In Fig.~\ref{fig:trajectory-transitions}(b) these are the arrows leaving the shaded region. 
Likewise, arrows pointing towards the shaded region illustrate transitions that end the excursion, corresponding to $W_{\!\!AB}$. Transitions that do not enter/leave the shaded region are part of  $\mathbb{W}_{\!\!B}$.
Since the excursions begin/end with both dots empty, transitions related to the double occupancy must always be inside $B$. Hence no $\tilde{f}_a$ appears in $W_{\!\!AB}$ and $W_{\!\!BA}$.
The $B$ block reads $\mathbb{W}_{\!\!B} = W_B - \Gamma_B$, where
\begin{equation}
\begin{aligned}
W_{\!\!B} &=
\begin{pmatrix}
\label{W-matrix-no-coulomb-no-diag}
0 & g_{\rm eff}& \gamma (1-\tilde{f}_R)\\
g_{\rm eff} & 0  & \gamma (1-\tilde{f}_L)\\
\gamma \tilde{f}_R & \gamma \tilde{f}_L & 0 \\
\end{pmatrix},
\\[0.2cm]
\Gamma_B &= \textrm{Diag}\big(\gamma(1-f_L) + g_{\rm eff} + \gamma \tilde{f}_R, \ \gamma(1-f_R) + g_{\rm eff} + \gamma \tilde{f}_L, \\[0.2cm]
&\qquad\qquad \   \gamma (1-\tilde{f}_R) +  \gamma (1-\tilde{f}_L)\big).
\end{aligned}
\end{equation}
Note that $\mathbb{W}_{\!\!B}$ is not a stochastic matrix because its columns do not add up to zero (rather, it is substochastic).

The steady state acquires a similar block decomposition structure
\begin{equation}\label{pss_blocks}
    \ket{p^{\rm ss}} = \begin{pmatrix}
        \ket{p_A^{\rm ss}}\\[0.2cm]
        \ket{p_B^{\rm ss}}
    \end{pmatrix}.
\end{equation}
We define the row vector $\bra{1} = (1,1,1,1)$, so that normalization of the probabilities can be written as $\bra{1}p\rangle = \sum_x p_x =1$. 
This vector is also similarly split in blocks as  $\bra{ 1} = ( \bra{ 1_A} ~~~\bra{1_B})$.

\subsection{Counting observables, excursion durations, and their distributions}

The notion of stochastic excursions has been around for many decades~\cite{Chung1976, Takacs1991}. However, much of the focus has been on quantities related to excursion times~\cite{Yao1985, Rubino1989, Kook1993, Takacs1974, Garrahan2017, Bebon2024, Roldan2019}. In fact, in that regard the problem is identical to a first passage time problem. 
The new feature of the formalism put forward in~\cite{Fiusa2025, Fiusa2025a} is the ability to augment that to also account for the statistics of counting observables. 
The basic idea is that we can count how many transitions occur during a single excursion, and we can use that to construct physical observables. 
Consider an individual excursion and let $\hat{N}_{xy}$ denote a random variable that counts how many times the transition $y \to x$ was observed in that single excursion (here and henceforth we use hats to denote random variables). 
A linear counting observable is a random variable defined as a linear combination of $\hat{N}_{xy}$, with arbitrary weights $\nu_{xy}$~\cite{landi2024a}:
\begin{equation}\label{counting_observable}
    \hat{Q} = \sum_{x,y} \nu_{xy} \hat{N}_{xy}.
\end{equation}
The rationale behind a counting observable is that physical quantities, such as the electric current, are related to specific transitions. For instance, the transition $\ket{00} \to \ket{10}$ must be associated with an electron tunneling from the left bath to the left dot. 
With the appropriate choice of weights $\nu_{xy}$ we can therefore construct any physical observable of interest. 
As an example, the matrix $\nu$ for the particle current flowing to the right reservoir is given in Eq.~\eqref{nu_current} below. 
Notice that this matrix is anti-symmetric, $\nu_{xy}=-\nu_{yx}$, a constraint that must be satisfied by any thermodynamic current~\cite{Wachtel2015}. 
Another example is the entropy production, whose $\nu$-matrix is given in Eq.~\eqref{nu_entropy}. 
Counting observables where $\nu$ is not anti-symmetric can also be of physical interest. The prime example is the dynamical activity, which counts how many transitions occur in a given excursion. The corresponding $\nu$ matrix is shown in Eq.~\eqref{nu_activity} below.

There are two relevant timescales in a stochastic excursion.
First, we define the excursion duration $\hat{T}$, as the time it takes for an excursion to take place, starting from the first time we observe $A\to B$ and ending when we first observe $B\to A$. 
After the excursion ends the system will spend a random residence time in region $A$, before another excursion begins. 
We denote this residence time as $\hat{\tau}$. 
When region $A$ has a single state, as is the case here, 
this will be an exponentially distributed random variable with mean residence time $E(\hat{\tau})= \Gamma_x^{-1}$ where, for us $x = \ket{00}$. 
Excursion times and residence times always alternate: after one excursion duration, there is a residence time, which is then followed by another excursion duration, and so on.
Compounded together, this therefore defines a stochastic cycle, with total cycle time 
\begin{equation}\label{cycle_time_random_variable}
    \hat{T}^{\rm cyc} = \hat{T} + \hat{\tau}.
\end{equation}
A cycle is then comprised of an excursion followed by a residency in $A$. 

If region $A$ has a single state, which is the case we deal with here, various simplifications occur. 
First,  $\hat{T}$ and $\hat{\tau}$ become statistically independent. 
Second, the entire stochastic trajectory becomes a  renewal process~\cite{cox1967renewal,ross2007introduction} (also called a Markovian event~\cite{Jann2023, JannPRX2022}).
This means that the statistics of a given cycle (both counting observables and cycle times) are statistically independent of previous cycles.  
The average cycle duration is 
\begin{equation}
\label{mu}
    \mu := E( \hat{T}^{\rm cyc}) = E(\hat{T}) + \dfrac{1}{\Gamma_x}.
\end{equation}
Since $\hat{T}$ and $\hat{\tau}$ are independent, the  variance of the total cycle time is just the sum of the variances: 
\begin{equation}
\label{delta2}
    \Delta^2 :=\textrm{var}(\hat{T}^{\rm cyc}) =   \textrm{var}(\hat{T}^{})+\dfrac{1}{\Gamma_x^2}.
\end{equation}

Among the main results of Refs.~\cite{Fiusa2025, Fiusa2025a}, a method to calculate the joint statistics of both counting observables $\hat{Q}$ and cycle times was developed. 
Consider a counting observable $\hat{Q}$ defined by weights $\nu_{xy}$. 
Define the tilted matrix 
\begin{equation}\label{tilted_transition_matrix}
    (\mathbb{W}_\xi)_{xy} =
    \begin{cases}
    W_{xy} e^{-i \nu_{xy} \xi} & x\neq y
    \\
    -\Gamma_x & x = y,
    \end{cases}
\end{equation}
where $\xi$ is an auxiliary variable called a counting field. 
The tilting is similarly applied to the individual blocks, $\mathbb{W}_{A\xi}$, $W_{AB\xi}$, etc. 
Let  $\ket{x_A} = (1,0,0,0)^{\rm T}$ denote a column vector, and $\bra{x_A} = (1,0,0,0)$ the corresponding row vector. 
The joint probability distribution that the counting observable  $\hat{Q}$ takes on a value $q$ and the excursion time $\hat{T}$ takes on a value $t$ is then given by~\cite{Fiusa2025, Fiusa2025a}
\begin{equation}\label{Pqt}
    P({q},t) =  \int\limits_{-\infty}^\infty \frac{d\xi}{2\pi} \frac{
    \bra{x_A} W_{AB{\xi}} e^{\mathbb{W}_{\!B{\xi}}t} W_{BA{\xi}} \ket{x_A} e^{i q\xi}}{
    \bra{x_A} W_{AB} (-\mathbb{W}_B^{-1}) W_{BA}\ket{x_A}
    }.
\end{equation}
The intuition behind result~\eqref{Pqt} is as follows. 
The presence of the counting field $\xi$ in the tilted matrices counts specific transitions. 
The first transition $A \to B$ is accounted for in $W_{BA\xi}$. Then all possible transitions within $B$ are recorded using $\exp(\mathbb{W}_{\!B{\xi}}t)$. Finally, the transition $B \to A$ that ends the excursion is recorded through $W_{AB \xi}$. 
Finally, the integration over $\xi$ then brings us from the characteristic function to the probability distribution.

The result in Eq.~\eqref{Pqt} contains all the information about the joint statistics of $\hat{Q}$ and $\hat{T}$. 
And since $\hat{T}$ and $\hat{\tau}$ are statistically independent, we can also readily compute the joint statistics of $\hat{Q}$ and $\hat{T}^{\rm cyc}$. 
Marginalizing $P(q,t)$ over $t$ yields the distribution of $\hat{Q}$. 
And likewise marginalizing over $q$ yields the distribution of $\hat{T}$. 
The full distribution $P(q,t)$ is needed to compute the  correlation between $\hat{Q}$ and $\hat{T}$. 
If one is only interested in statistical moments, and not the full probability distribution,  simple-to-use formulas are provided in Ref.~\cite{Fiusa2025a}.

\subsection{Connection to steady-state current and noise}
\label{subsec:current-and-noise}

Excursion-related quantities (described by $P(q,t)$) can also be related to more familiar quantities pertaining to the steady-state.
Consider a very long stochastic trajectory and define $\hat{\mathcal{N}}_{xy}(t)$ as the random variable counting how many transitions from $y\to x$ are observed in the interval $[0,t]$, where $t$ is assumed to be very large.
Counting observables for long trajectories are defined in a similar way: 
\begin{equation}
    \mathcal{Q}(t) = \sum_{x,y} \nu_{xy} \hat{\mathcal{N}}_{xy}(t).
\end{equation}
The average steady-state current is then defined as 
\begin{equation}
    J_Q = \frac{dE(\mathcal{Q})}{dt}.
\end{equation}
And it can be written in terms of steady-state quantities as~\cite{Esposito2007} 
\begin{equation}
    J_Q = \sum_{x,y} \nu_{xy} W_{xy} p_y^{\rm ss} \equiv \braket{1|\mathcal{J}|p_{\rm ss}},
\end{equation}
where $\mathcal{J}_{xy} := \nu_{xy} W_{xy}$. 
As shown in~\cite{Fiusa2025}, the connection with  excursion-related quantities is intuitively given by 
\begin{equation}
\label{J-exc}
    J_Q = \dfrac{E(\hat{Q})}{\mu},
\end{equation}
where $E(\hat{Q})$ is the average of the counting observable within a single excursion and $\mu$ is the average cycle time in Eq.~\eqref{mu}.
The right-hand side therefore only contains quantities associated to individual excursions, whereas the left-hand side is a steady-state quantity. 

We can also find a similar relation for the steady-state current fluctuations, defined in terms of the noise (also called the scaled variance, or the diffusion coefficient)~\cite{landi2024a,Esposito2007}.
It is formally defined as 
\begin{equation}\label{noise_def}
    D_Q = \frac{d}{dt}{\rm var}(\mathcal{Q}(t)),
\end{equation}
and can be computed in practice using~\cite{landi2024a,Esposito2007}
\begin{equation}
    D_Q = \sum_{x,y} \nu_{xy}^2 W_{xy}p_{y}^{\rm ss} + 2 \int_0^\infty dt \Big(\braket{1|\mathcal{J} e^{\mathbb{W}t}\mathcal{J}|p_{\rm ss}}-J_Q^2\Big) .
\end{equation}
As shown in~\cite{Fiusa2025}, the connection between the steady-state noise and excursion-related quantities is given by
\begin{equation}
\begin{aligned}
\label{D-exc}
     D_Q &= \frac{{\rm var}(\hat{Q})}{\mu} + \frac{\Delta^2}{\mu^3}E(\hat{Q})^2 - \frac{2E(\hat{Q})}{\mu^2}{\rm cov}(\hat{Q},\hat{{T}})\\[0.2cm]
     &=: D_1 + D_2 + D_3,
\end{aligned}
\end{equation}where $\mu$ and $\Delta^2$ were defined in Eqs.~\eqref{mu} and \eqref{delta2}. 
The steady-state fluctuations therefore have three distinct contributions, with clear interpretations. 
The first term describes the fluctuations of the counting observable alone. The second describes fluctuations of the excursion time. And the third describes correlations between the counting observable and the excursion time. 
Results \eqref{J-exc} and~\eqref{D-exc} can be used to characterize \emph{any} counting observable of interest in the double quantum dot system.

\section{Results}
\label{sec:results}

Throughout, we write formulas with $\hbar = k_B = 1$, so that $h = 2\pi$. Unless otherwise stated, numerical values are reported as ordinary frequencies in GHz: energies are divided by $h$ and rates by $2\pi$.
We assume the following parameters, in units of GHz: $g = 1~ (\sim4.1\ \mu \textnormal{eV})$, $\gamma = 0.63$, $k_B T = 2~(\sim 96 \ \textnormal{mK})$, $U = 10 ~(\sim41~\mu \textnormal{eV})$. 
Moreover, $V_g$ and $V_{sd}$ denote energies as we set $e = 1$ and absorb the gate lever arm, and following the units described here, they are also reported as $E/h$ in GHz.
A reported value of $1$ GHz thus corresponds to an energy unit of $h ×\times 1$ GHz $\approx 4.1$ $\mu$eV, a temperature unit of $48$ mK, and a time unit of $\approx 0.16$ ns.
It is convenient to write the transitions between the dots in terms of their effective coupling $\geff$, which under the values chosen here reads $\geff = 2g^2/\gamma \simeq 3.2$.
For context, typical tunnel couplings $g$ in gate-defined semiconductor quantum dots range from a few to tens of $\mu$eV, typical electron temperatures are about 100 mK~\cite{dicarlo2004differential}, and typical dot-reservoir coupling strengths $\gamma$ can vary from 0 to larger than the interdot coupling. Typical inter-dot capacitive couplings are of order 0.01-1~meV~\cite{dicarlo2004differential}.    
We consider the case where $U\to \infty$ in Section~\ref{subsec:suc-fail-dis}
and in Section~\ref{subsec:tur}, where we discuss the effect of the blockade in the Fano factor. When  $U\to \infty$, simple analytical formulas can be obtained for all quantities we discuss next. We delegate this discussion and the formulas to Appendix~\ref{app:analytical}.

\subsection{Cycle duration and fluctuations}
\label{subsec:excursion-duration}

We begin by analyzing the statistics of the cycle times $\hat{T}^{\rm cyc}$. 
As defined in Eq.~\eqref{cycle_time_random_variable}, this is comprised of an excursion time $\hat{T}$ and a residence time $\hat{\tau}$ in $A$.
Excursions begin either with $\ket{00} \to \ket{10}$ or $\ket{00} \to \ket{01}$ transitions, which are described by rates $\gamma f_L$ and $\gamma f_R$, respectively. Likewise, they end with the reverse transitions which have rates $\gamma (1-f_L)$ and $\gamma (1-f_R)$. 
Even for moderate temperatures, the Fermi functions change sharply from 0 to 1, as a function of the argument. 
For this reason, the total cycle time $\hat{T}^{\rm cyc}$ will be essentially dominated by either $\hat{T}$ or $\hat{\tau}$.
When the Fermi functions approach unity, the residence time will be small. So most of the cycle time will be with the system undergoing an excursion. 
Conversely, when the Fermi functions approach zero, the rates for starting the excursion become small, and the system will spend a significant amount of time residing in $\ket{00}$. 

\begin{figure}
    \centering
    \includegraphics[width=1.0\linewidth]{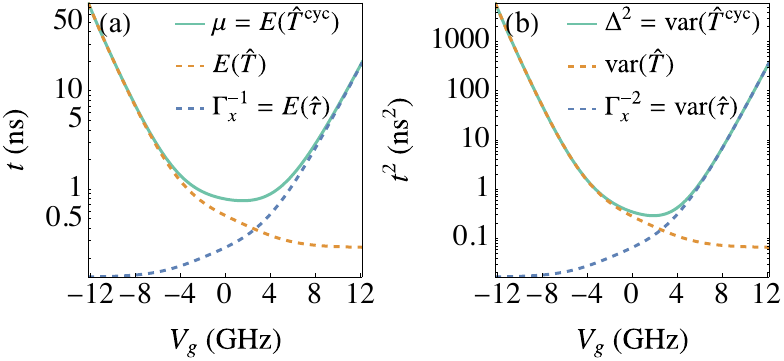}
    \caption{(a) Average cycle, excursion, and residence times in log scale, in ns, as a function of gate voltage $V_g$. 
    (b) Variance of cycle, excursion, and residence times in log scale, in ns$^2$, as a function of gate voltage $V_g$. For both plots, we fixed the following parameters (in GHz): $\geff = 3.2$, $\gamma = 0.63$, $k_BT = 2$, $U = 10$, and $V_{sd} = 7$.}
    \label{fig:avg-var-times}
\end{figure}

In Fig.~\ref{fig:avg-var-times} we plot the average and variance of the cycle times, excursion durations, and residence times for a fixed value of source-drain bias $V_{sd}$.
In the region of $V_g <0$, residence times are quite short while excursion durations are long, and the noise is dominated by fluctuations in excursion duration.
This corresponds to the regime where $f_L, ~f_R\to 1$ for very low temperatures.
Conversely, for $V_g > 0$, the system now spends a long time in $A$ (notice the log scale).
This corresponds to the regime where $f_L,~f_R \to 0$ for very low temperatures.

\subsection{Current and noise}
\label{subsec:current-and-noise}

\subsubsection{Particle current}
\label{subsection:transport}
\begin{figure}
    \centering
    \includegraphics[width=1.0\linewidth]{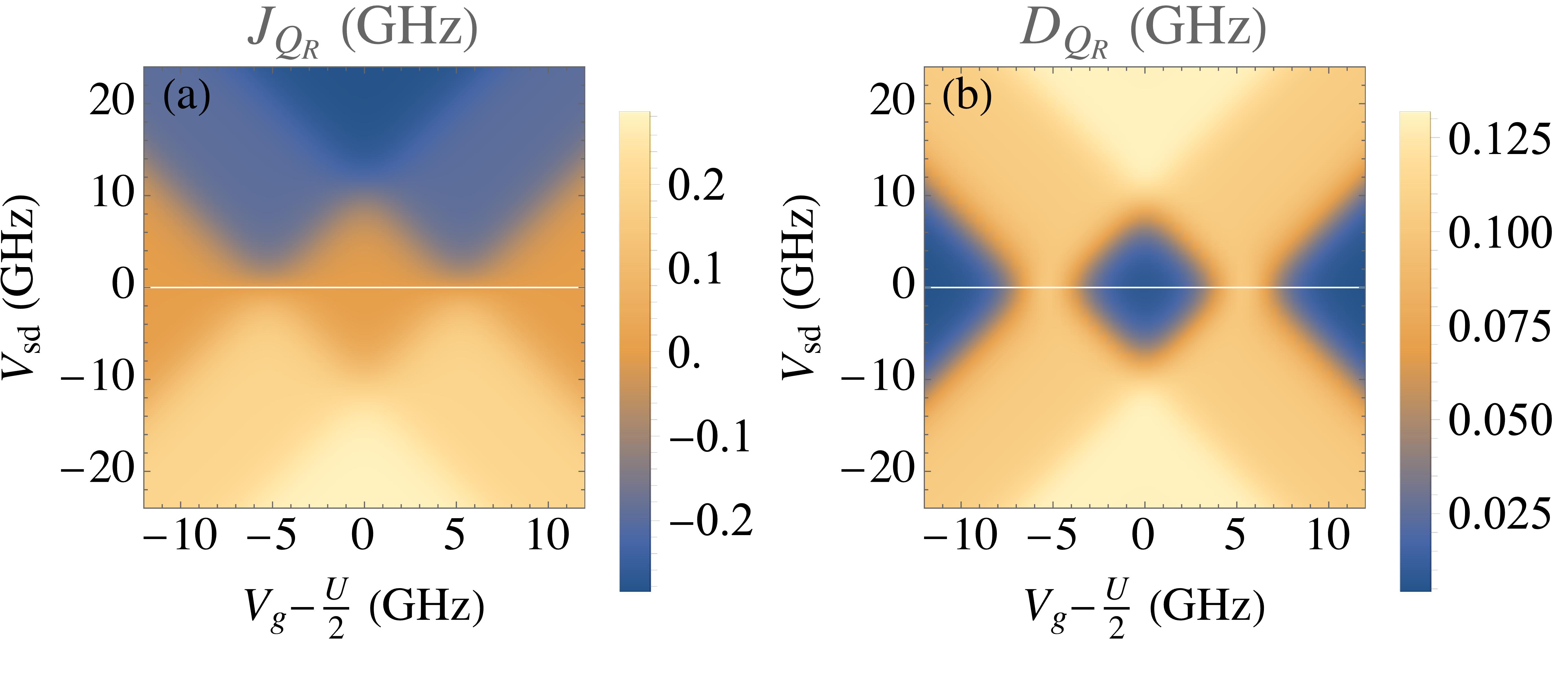}
    \caption{(a) Current and (b) Diffusion coefficient of the transport observable $\hat{Q}_R$ visualized in the Coulomb diamond $V_g \times V_{sd}$. For both plots we considered (in units of GHz) $\geff = 3.2$, $\gamma = 0.63$, $k_BT = 1$, $U = 10$.}
    \label{fig:transport-current-D}
\end{figure}

Recall that counting observables can be constructed by introducing a weight matrix $\nu$, as in Eq.~\eqref{counting_observable}. 
In this section we study the particle current to the right bath, which is  constructed with weights
\begin{equation}\label{nu_current}
   \nu_{Q_R} = \begin{pmatrix}
0 & 0 & 1&0\\
0 & 0 & 0&1\\
-1 & 0 & 0 &0\\ 
0 & -1 & 0 &0\\ 
\end{pmatrix}.
\end{equation}
Note that it suffices to study the particle current of only one of the two reservoirs. Indeed, we show in Appendix~\ref{app:conservation} that within a single excursion the two transport observables obey
\begin{equation}
\hat{Q}_L = -\hat{Q}_R,    
\end{equation}
not merely on average, but as an identity between random variables. 
This conservation law relies directly on the structure of excursions.
The reason is that an excursion begins and ends in the state $\ket{00}$, so every excitation entering the system must leave it before the excursion ends. 
This is in contrast with the corresponding statement for long trajectories, where the relation need not to hold for a given trajectory snippet, and all that can be said is about averages.
The excursion filtering therefore sharpens this relation into an exact identity at the level of individual realizations.

A common way of visualizing transport in double quantum dots is through density plots in terms of $V_g$ and $V_{sd}$. 
This leads to the characteristic  \emph{Coulomb diamond} shape~\footnote{For all Coulomb diamond plots, we rescale $V_g \to V_g - U/2$. We can always add or subtract a constant to $V_g$ because the origin of the gate voltage is arbitrary in the experiment. The reason behind this rescaling is to have the center at the diamond in $V_g = 0$. Otherwise we would have it in $V_g = U/2$.} in Fig.~\ref{fig:transport-current-D}, where we plot the average particle current $J_{Q_{R}}$ and the noise $D_{Q_R}$.
The average current is maximal whenever the energy of each dot $V_g$ is small but the source-drain bias $V_{sd}$ is large. 
This regime is characterized by two features. 
First, the difference between Fermi energies, where injections from the left and extractions from the right baths are more likely than their reverse counterparts. 
The second feature is the interaction between the dots, which effectively is what allows transport to happen. 
The current suppression at large $|V_g|$ comes from the Fermi functions, which dictate extractions and injections.

There are three regions where the current is vanishingly small.
In the left-most region, where $V_g$ is large and negative, and $V_{sd}$ is close to zero, the transition rates for excursions ending are very small, which hinders the transport properties. 
This is already expected from the analysis of cycle times in the previous section. 
Namely, the left corner for small $V_{sd}$ is excursion duration-dominated, which means that excursions last for a long time, making  transport highly inefficient. 
Conversely, the right-most region, where $V_g$ is large and  positive, and $V_{sd}$ close to zero, is residence time-dominated, which means that transitions that begin excursions are highly unlikely. 
As a consequence, this similarly results in a very small current. 

The middle region (i.e., inside the Coulomb diamond), provides an interesting case study.
In this region, transition rates of extractions and injections are roughly the same.
The reason why the current is small in this region is therefore not dominated by the cycle durations. Instead, it is because excursions with positive and negative transport --- corresponding, respectively, to clockwise and counter-clockwise cycles in the diagram of Fig.~\ref{fig:trajectory-transitions}(b) --- are on average equally likely.
Finally, as for the noise (Fig.~\ref{fig:transport-current-D}(b)), we see that higher currents also come at a higher noise cost.

\subsubsection{Dynamical activity}\label{sec:dynamical-activity}

The second observable we explore in this study is the dynamical activity, which has recently come to prominence due to its appearance in uncertainty relations~\cite{Di_Terlizzi2018, Saito2008, Gingrich2016, Hasegawa2019a, Hasegawa2019b, Van_Vu2019, Van_Vu2020, Vo2020b, Vo2022, Van_Vu2023, vanvu2025universalprecisionlimitsgeneral}.
In the steady-state, the dynamical activity is defined as the average number of transitions per unit time.
In the context of excursions, we can therefore study the statistics of the number of transitions \emph{per} excursion.
We denote this counting observable by $\hat{\mathcal{A}}$.
It can be constructed using 
\begin{equation}\label{nu_activity}
   \nu_{\mathcal{A}} = \begin{pmatrix}
0 & 1 & 1 &1\\
1 & 0 & 1 &1\\
1 & 1 & 0 &1\\
1 & 1 & 1 &0\\
\end{pmatrix}.
\end{equation}
Because of the properties of excursions, the number of transitions is always greater or equal to $2$.
The connection with the steady-state dynamical activity is given by Eq.~\eqref{J-exc}:
\begin{equation}
\label{eq:JA}
    J_{\mathcal{A}} = \frac{E(\hat{\mathcal{A}})}{\mu} = \sum_{x,y} W_{xy} p_y^{\rm ss} =\sum_{x} \Gamma_x p_x^{\rm ss}.
\end{equation}
In Fig.~\ref{fig:J-entropy-A}(a) we plot $J_{\mathcal{A}}$ as a function of $V_g$ and $V_{sd}$. 
When contrasting the activity with the particle current [see Fig.~\ref{fig:transport-current-D}], we see that the regions in the far right and far left corner have little excursion activities and no transport.
In contrast, the Coulomb diamond in the middle is the region with the most activity, despite having a vanishing current. 
This is consistent with the aforementioned fact that, inside the diamond, excursions happen frequently, but positive and negative transport excursions tend to happen with roughly equal probability.
In fact, inside the diamond the dominating transition is the effective coupling between the dots $g_{\rm eff}$, which means that once an excursion starts ($A\to B$), the next transitions are quite likely to be the hopping between the dots.
Moreover, in the central diamond, the rates are such that $\gamma f_L \simeq \gamma f_R$, which means excursions are approximately equally likely to end up in positive and negative transport. 
The net effect is therefore a vanishing transport but a large number of transitions within each individual excursion (i.e. many in-between hoppings).

\subsubsection{Entropy production}

\begin{figure}
    \centering
    \includegraphics[width=1.0\linewidth]{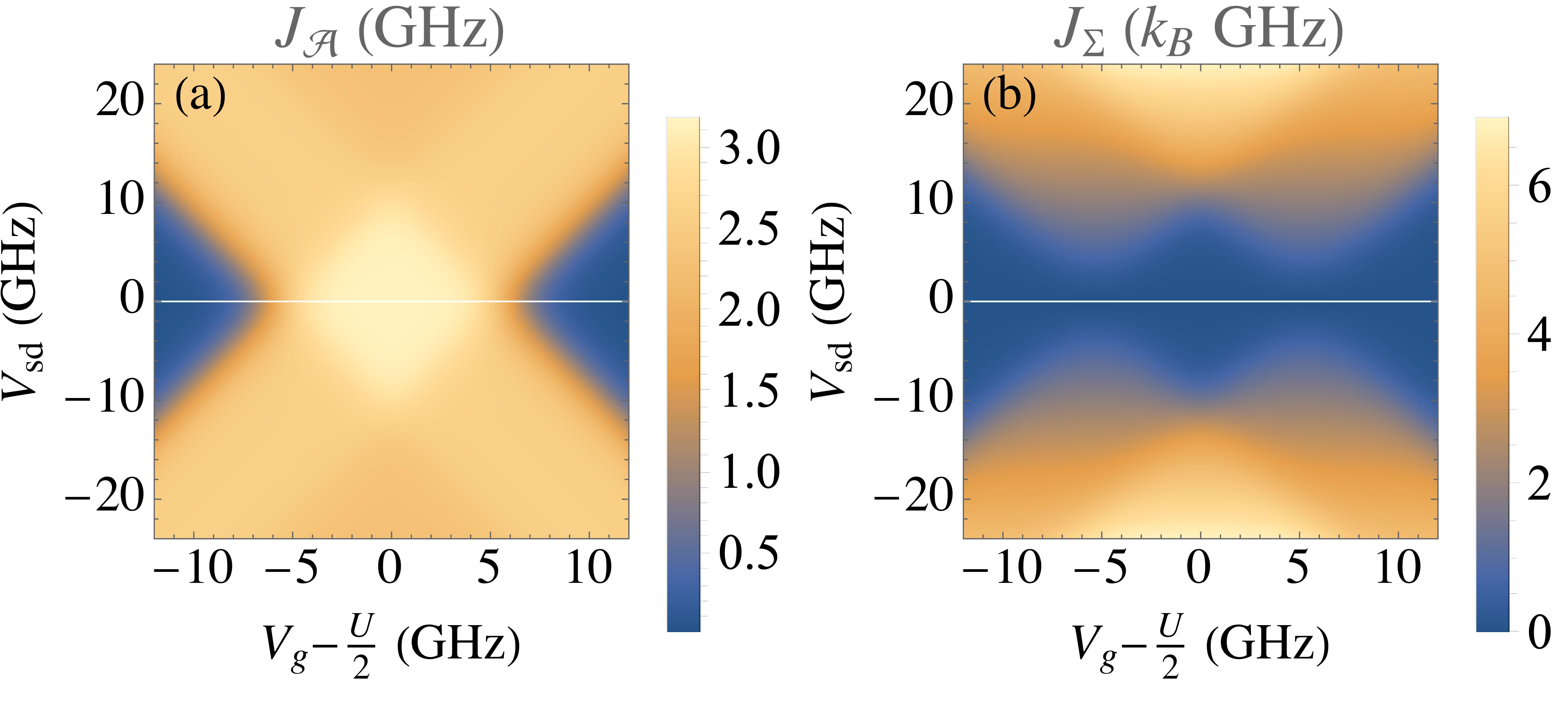}
    \caption{Current of (a) Dynamical activity $\hat{\mathcal{A}}$  and (b) Entropy production $\hat{\Sigma}$  visualized in the Coulomb diamond $V_g \times V_{sd}$. For both plots we considered (in units of GHz) $\geff = 3.2$, $\gamma = 0.63$, $k_BT = 1$, $U = 10$.
    }
    \label{fig:J-entropy-A}
\end{figure}

The third and last counting observable we consider is the entropy production. 
This observable is truly the cornerstone of stochastic thermodynamics, as it signals how far from equilibrium a physical system is. 
Inference of counting observables such as entropy production in hidden/coarse grained settings is currently an active area~\cite{rahavFluctuationRelationsCoarsegraining2007, boEntropyProductionStochastic2014, biskerHierarchicalBoundsEntropy2017, yuDissipationLimitedResolutions2024a, harunariWhatLearnFew2022, espositoStochasticThermodynamicsCoarse2012, ehrichTightestBoundHidden2021, blomMilestoningEstimatorsDissipation2024, liangThermodynamicBoundsSymmetry2024, Tan2021, Jann2023, JannaPRR2024, Maier2025, JannPRX2022, seifertUniversalBoundsEntropy2025, harunariUncoveringNonequilibriumUnresolved2024, ertelEstimatorEntropyProduction2024, skinnerEstimatingEntropyProduction2021, martinezInferringBrokenDetailed2019}.
Moreover, akin to the dynamical activity, the so-called thermodynamic uncertainty relation shows that the entropy production bounds the precision of any thermodynamic current~\cite{Barato2015, Gingrich2016,Hasegawa2019b, Hasegawa2019a, Horowitz2019, Vo2022, Gingrich2016, liQuantifyingDissipationUsing2019, Hasegawa2019a, Hasegawa2019b, Van_Vu2019, Van_Vu2020, Vo2020b, Vo2022, Van_Vu2023, vanvu2025universalprecisionlimitsgeneral}.
The entropy production is constructed by using weights 
 \begin{equation}\label{nu_entropy}
   \nu_{\Sigma} = \begin{pmatrix}
0 & \log\left(\frac{1-f_L}{f_L}\right) & \log\left(\frac{1-f_R}{f_R}\right) &0\\
\log\left(\frac{f_L}{1-f_L}\right) & 0 & 0 &\log\left(\frac{1-\tilde{f}_R}{\tilde{f}_R}\right) \\
\log\left(\frac{f_R}{1-f_R}\right) & 0 & 0 &\log\left(\frac{1-\tilde{f}_L}{\tilde{f}_L}\right)\\
0 & \log\left(\frac{\tilde{f}_R}{1-\tilde{f}_R}\right) & \log\left(\frac{\tilde{f}_L}{1-\tilde{f}_L}\right) & 0\\
\end{pmatrix}
\end{equation}
in Eq.~\eqref{counting_observable}.
Similarly to the transport observable, entropy production is also a thermodynamic current in the sense that $\nu$ is anti-symmetric. 
This dictates many of its properties.
In particular, we show in Appendix~\ref{app:conservation} that the average entropy production rate is related to the average particle current by
\begin{equation}
    J_\Sigma = \log\left( \frac{f_L(1-f_R)}{f_R(1-f_L)}
    \right) J_{Q_R}.
\end{equation}
A similar relation also holds true for the noise, see Eq.~\eqref{eq:currents} and the derivation in Appendix~\ref{app:conservation}.
We plot the average entropy production rate in Fig.~\ref{fig:J-entropy-A}(b). 
As expected,  it closely follows the particle current (Fig.~\ref{fig:transport-current-D}).
One particularly important difference is the fact that $E(\hat{\Sigma} ) \geq 0$, in contrast with the transport current that can be negative.
Comparing the entropy production current with the dynamical activity also provides some insight into the nature of the hopping between the dots. 
In the transition matrix~\eqref{W-matrix-no-coulomb-no-diag}, the only contribution that is unchanged if we transpose it, is $g_{\rm eff}$. 
This is a consequence of the fact that $g_{\rm eff}$ is a unitary contribution, associated to the tunneling from one dot to the other. 
Hence, it follows that the transitions $\ket{10} \leftrightarrow \ket{01}$ produce no entropy. 
This explains why $J_\Sigma \simeq 0$ inside the Coulomb diamond in Fig.~\ref{fig:J-entropy-A}(b) since, as already discussed above, this region is dominated precisely by the $\ket{10} \leftrightarrow \ket{01}$ transitions.
Similarly to the current, the far left corner produces no entropy because excursions become too long, and the far right corner produces no entropy because residence times become too long and there is too much time in-between excursions.

\subsection{Steady-state populations }
\label{subsec:pop}

\begin{figure}
    \centering
    \includegraphics[width=1.0\linewidth]{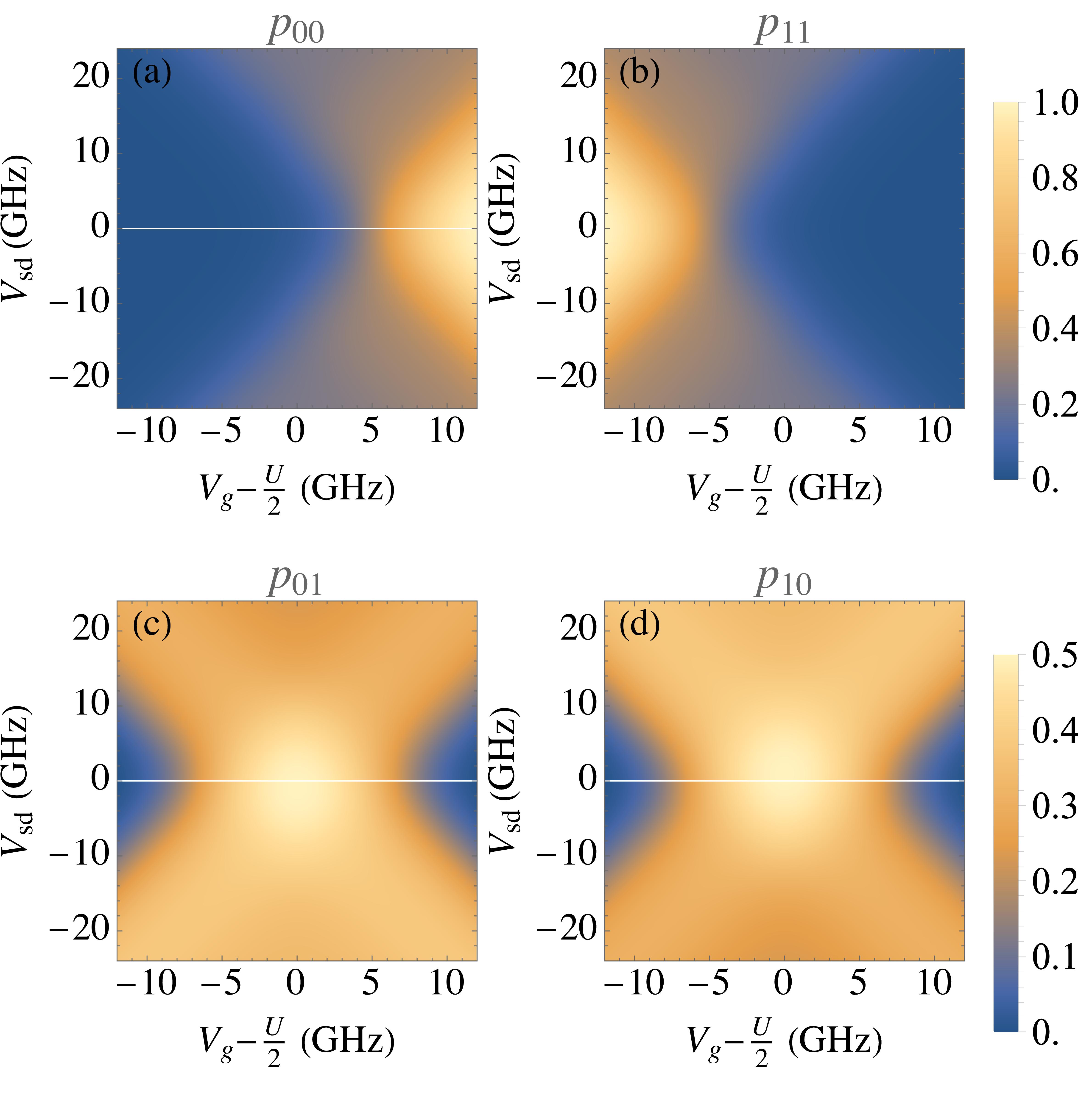}
    \caption{Probabilities for the distribution of populations visualized in the Coulomb diamond.
    (a) $p_{00}$ represents the probability that both dots are empty, (b) $p_{11}$ represents double occupancy, (c) $p_{01}$ represents that the right dot is occupied and (d) $p_{10}$ represents the left dot is occupied.
    The normalization is such that $p_{00} + p_{01} + p_{10} + p_{11} = 1$ for the full parameter range.
    Here we considered (in units of GHz) $\geff = 3.2$, $\gamma = 0.63$, $k_BT = 2$, and $U = 10$.}
    \label{fig:population-heat-map}
\end{figure}

We briefly summarize here the behavior of the steady-state populations $p_x^{\rm ss}$, obtained by solving $\mathbb{W}\ket{p^{\rm ss}} = 0$.
The results are shown in Fig.~\ref{fig:population-heat-map}.
In the far right corner, we see that the system is almost always in the state $\ket{00}$. 
This precisely corresponds to the intuition we had with excursions where in this regime, residence times are by far the dominating timescale [see Fig.~\ref{fig:avg-var-times}(a)].
Long residence times means that the system spends long times in state $\ket{00}$ before starting an excursion, and this is confirmed by assessing the actual distribution of populations.
Likewise, in the far left corner, excursion durations are the dominant timescale [see Fig.~\ref{fig:avg-var-times}(a)], so excursions in general take very long to end. This corresponds to a long time visiting the double occupation state $\ket{11}$, because the end of an excursion requires the system to revisit the state $\ket{00}$.
Once more, analyzing the distribution of populations also corroborates our results for the current [see Fig.~\ref{fig:transport-current-D}(a)]. Regions where excursions either take too long to end or the residence time is too long, have little transport. Accordingly, they correspond to regions  where the states $\ket{00}$ and $\ket{11}$ are the dominating ones.

One further insight that can be extracted from the populations is that inside the Coulomb diamond the system spends roughly equal amounts of time in  $\ket{01}$ and $\ket{10}$.
This has several consequences.
As observed for the transport current, this means that there is no net transport within the region and no entropy production.
For the dynamical activity current, however, there is maximum activity. 
This corresponds to our intuition that in the middle diamond the excitation keeps hopping between the two dots, and as a consequence, excursions on average have more jumps than in other regions, because the transition $g_{\rm eff}$ is the dominating one.

\begin{figure}
    \centering
    \includegraphics[width=0.6\linewidth]{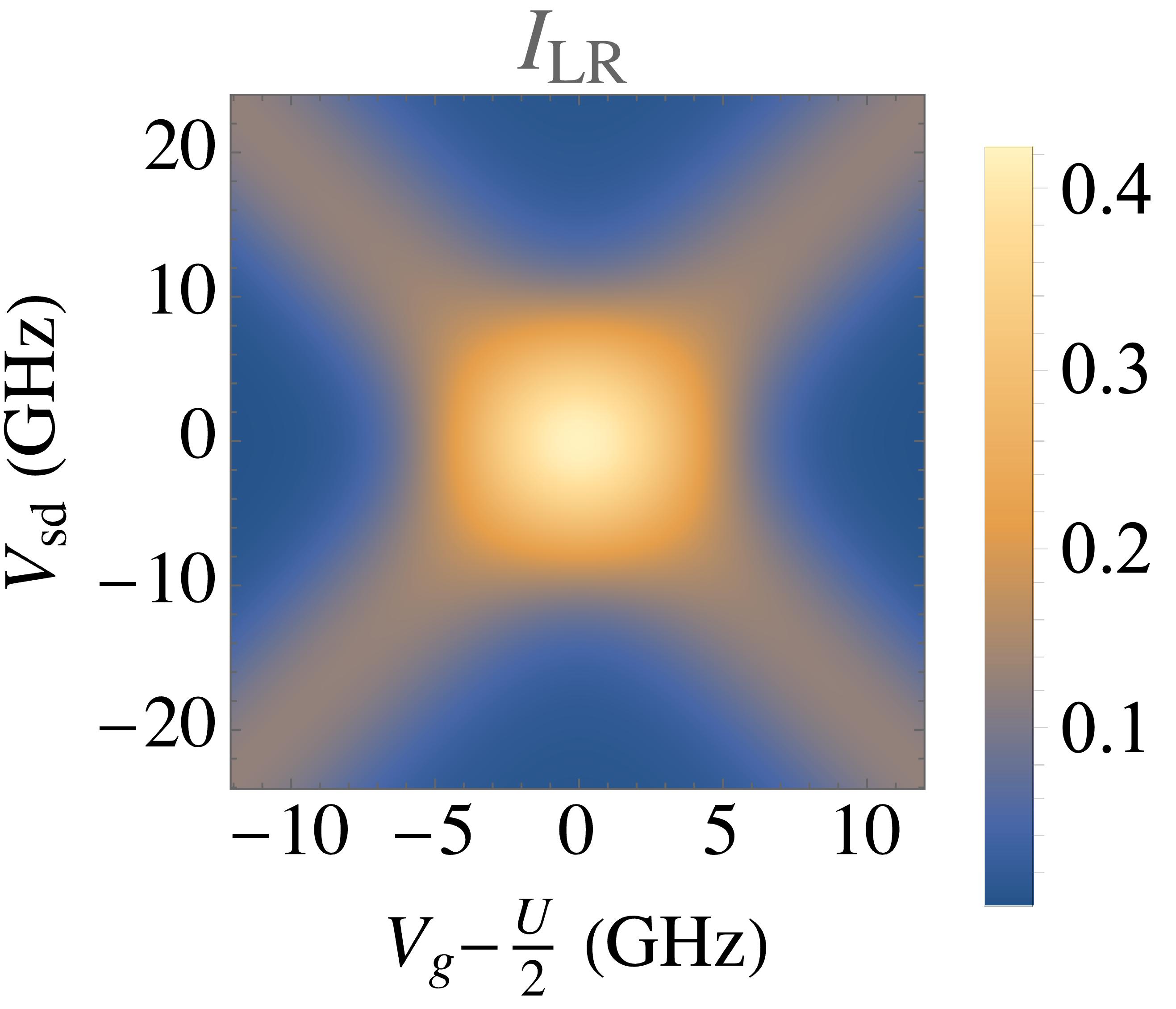}
    \caption{Mutual information between the occupation of the two dots, Eq.~\eqref{eq:mutual-info}, visualized in the Coulomb diamond. 
    Here we considered (in units of GHz) $\geff = 3.2$, $\gamma = 0.63$, $k_BT = 2$, and $U = 10$.}
    \label{fig:mutual-info}
\end{figure}

In Fig.~\ref{fig:mutual-info} we plot the mutual information between the two dots, which reads
\begin{equation}
\label{eq:mutual-info}
I_{\rm LR} = \sum_{n_L, n_R = 0,1}
  p^{\rm ss}_{n_L n_R}\,
  \log\!\left[\frac{p^{\rm ss}_{n_L n_R}}{p_{n_L}\, p_{n_R}}\right],
\end{equation}where $n_L, n_R \in \{0,1\}$ denotes the occupation of the left and right dot, and the steady state $p^{\rm ss}_{n_L n_R}$ is their joint distribution, with marginals $p_L := p^{\rm ss}_{10} + p^{\rm ss}_{11}$ and $p_R := p^{\rm ss}_{01} + p^{\rm ss}_{11}$ giving the probability that the left (right) dot is occupied, irrespective of the other.
This quantity vanishes when the two dots are statistically independent and is largest in the region dominated by the hopping between dots, where the system spends nearly all its time in $\ket{10}$ or $\ket{01}$ and the occupations are therefore strongly anticorrelated.

\subsection{Quantifying successes, failures, and disasters}\label{subsec:suc-fail-dis}

Through the excursions framework we can classify the probability distribution of outcomes of counting observables based on the possible excursion trajectories.
In this and in the very next subsections, we assume that the system is under the Coulomb blockade regime; i.e. $U \to \infty$. 
As a consequence, the particle current or any other thermodynamic current may have only three possible outcomes per excursion.
We focus on the former. 
The possibilities are then as follows. 
In a single excursion the particle current can be $+1$, which configures a successful transport where the excitation moved from the left reservoir and was dumped into the right.
We label this as a \emph{success}. 
Or the particle current can be $-1$. This corresponds to the exact opposite, and we thus label it as a \emph{disaster}. 
Finally, the particle current can be zero.
This corresponds to the case where the electron went from one of the reservoirs to the system, but then back to the same reservoir. 
We label this a \emph{failure}.

The full distribution of any counting observable is obtained by marginalizing Eq.~\eqref{Pqt} over $t$, which yields 
\begin{equation}\label{Pq}
    P({q}) =  \int\limits_{-\infty}^\infty \frac{d\xi}{2\pi} \frac{
    \langle x_A| W_{AB{\xi}}  (-\mathbb{W}_{B\xi}^{-1}) W_{BA{\xi}} \ket{x_A} e^{i q\xi}}{
    \langle x_A|W_{AB} (-\mathbb{W}_B^{-1}) W_{BA}\ket{x_A}
    }.
\end{equation}
For the particular case of the Coulomb blockade the support of the counting observable becomes discrete and this reduces to 
\begin{equation}
\begin{aligned}
\label{transport-current-expanded}
    P(q_R) &= \int_{-\infty}^{\infty} \dfrac{d 
\xi}{2\pi} \left(P_\textrm{suc}~e^{-i\xi} + P_\textrm{dis}~e^{i\xi}+P_\textrm{fail}\right) e^{i q_R \xi}\\[0.2cm]
&=P_{\rm suc} \delta(q_R -1) + P_{\rm fail} \delta(q_R-0)+P_{\rm dis} \delta(q_R +1),
\end{aligned}
\end{equation}
where
\begin{align}\label{eq:psuc}
    P_{\rm suc} &= \dfrac{\geff f_L (1-f_R)}{(f_L + f_R)\Theta },
    \\[0.2cm]\label{eq:pfail}
    P_{\rm fail}&=\dfrac{\geff \left[f_L (1-f_L) + f_R (1-f_R)\right] + \Lambda (f_L + f_R)}{(f_L + f_R)\Theta},
    \\[0.2cm]\label{eq:pdis}
     P_{\rm dis} &= \dfrac{\geff f_R(1-f_L)}{(f_L + f_R)\Theta}.
\end{align}
Here $\geff$ follows from Eq.~\eqref{eq:g_eff} and we defined $\Lambda := \gamma(1-f_L)(1-f_R)$ and $\Theta := \Lambda + \geff\,(2-f_L-f_R)$ for simplicity.
One can readily verify that $P_{\rm suc} + P_{\rm  fail} +P_{\rm dis} =1$.

\begin{figure}
    \centering
    \includegraphics[width=1.0\linewidth]{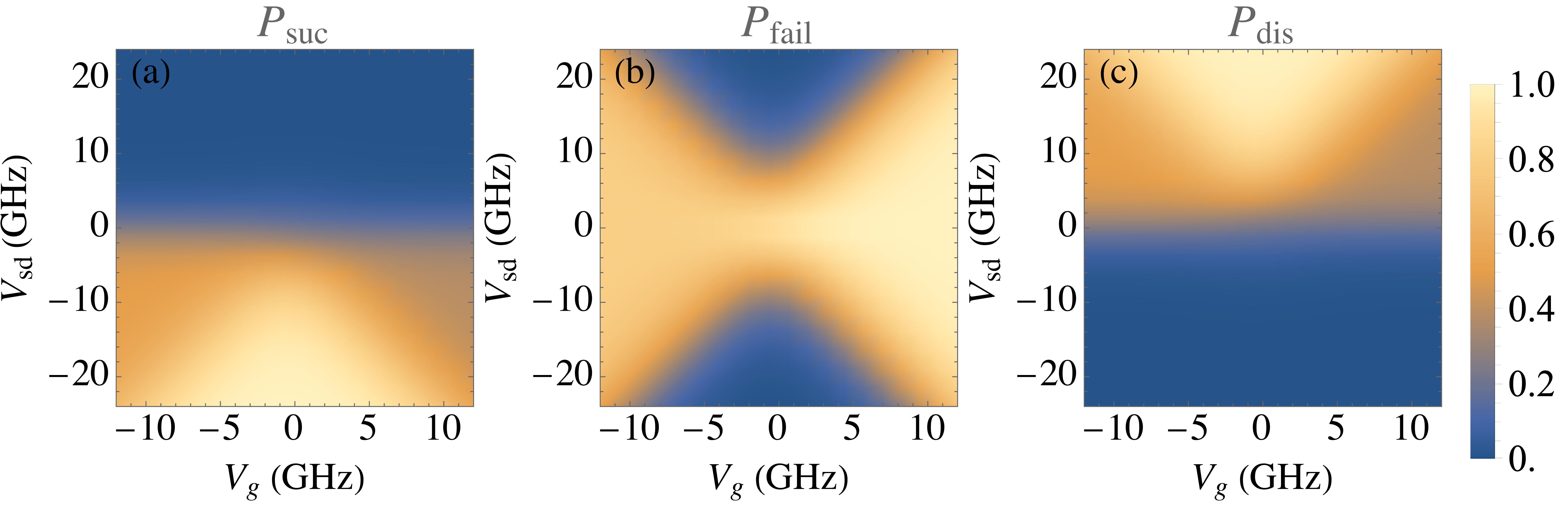}
    \caption{Probabilities for the transport counting observable within an excursion visualized in the Coulomb diamond under the blockade regime. $P_\textrm{suc}$ represents $P(\hat{Q}_{R}= 1)$, $P_\textrm{fail}$ represents $P(\hat{Q}_{R} = 0)$, and $P_\textrm{dis}$ represents $P(\hat{Q}_{R}= -1)$. 
    Here we considered (in units of GHz) $\geff = 3.2$, $\gamma = 0.63$, $k_BT = 2$.}
    \label{fig:zones-psuc-pfail-pdis}
\end{figure}

\begin{figure}
    \centering
    \includegraphics[width=1.0\linewidth]{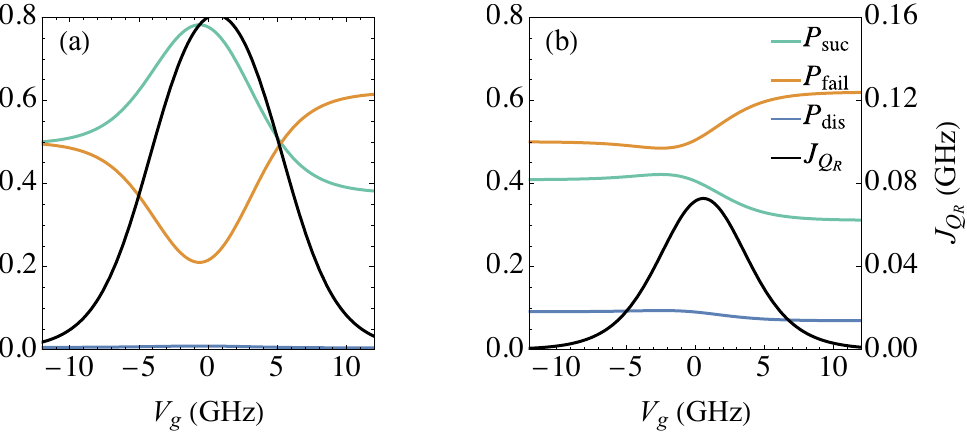}
    \caption{Probabilities for the transport counting observable within an excursion as a function of $V_g$. $P_\textrm{suc}$ represents $P(\hat{Q}_{R}= 1)$,  $P_\textrm{dis}$ represents $P(\hat{Q}_{R}= -1)$, and  $P_\textrm{fail}$ represents $P(\hat{Q}_{R} = 0)$.
    For illustrative purposes, we also plot the transport current $J_{Q_R}$, whose values (in GHz) are illustrated in the right ticks.
    Here we consider in GHz (a) $V_{sd}=-9$ and (b) $V_{sd}=-3$. For other parameters we considered (in units of GHz) $\geff = 3.2$, $\gamma = 0.63$, $k_BT = 2$.
    }
    \label{fig:psuc-pfail-pdis}
\end{figure}

In Fig.~\ref{fig:zones-psuc-pfail-pdis} we show the probability distributions of each outcome in the $V_g/V_{sd}$ plane.
The connection between $P_{\rm suc}, P_{\rm fail}$ and $P_{\rm dis}$ is established by Eq.~\eqref{J-exc} which, for the particular case of the particle current, becomes 
\begin{equation}
    J_{Q_R} = \frac{P_{\rm suc} - P_{\rm dis}}{\mu},
\end{equation}
an expression that is only true in the Coulomb blockaded regime (since in this case $\hat{Q}_R = 1,-1,0$).
Regions where $P_\textrm{suc}$ is large in comparison to $P_\textrm{dis}$ correlate with regions of high positive current, and vice-versa.
However, notice how the average current contains no information on failures. 
Fig.~\ref{fig:psuc-pfail-pdis} plots the same quantities as a function of $V_g$ for two different values of $V_{sd}$.
For the sake of comparison, we also include the average particle current, whose values are illustrated in the right ticks of plot (b). 
Under the blockade regime, $E(\hat{Q}_R)$ is limited between $-1$ and $1$, and the average current $J_{Q_R}$ is positive in Figs.~\ref{fig:psuc-pfail-pdis}(a) and (b) because we chose a negative $V_{sd}$.   
This plot better illustrates the competition between successes and disasters, and highlights that fails are truly the dominating processes over a wide range of parameters and thus should be taken into account for useful transport protocols.

\subsection{Fano factor}
\label{subsec:tur}
We now contextualize our study into thermodynamics of precision. 
We begin by exploring the Fano factor, which effectively quantifies dispersion in a stochastic process. It may also be thought of as a ``normalized noise''. 
Typically it is easier to assess experimentally than the noise itself~\cite{Rotter2004,Lewenkopf2008}.
It is defined for any counting observable $\hat{Q}$ as
\begin{equation}\label{eq:fano}
    F_Q = \dfrac{D_Q}{J_Q},
\end{equation}
where $D_Q$ is the steady-state noise [Eq.~\eqref{D-exc}] and $J_Q$ is the average steady-state current [Eq.~\eqref{J-exc}] \footnote{Since the Fano factor depends on the average current, and for the transport observable we have a parameter range ($V_{sd} < 0$) where the average current is negative, we could potentially end up with a negative Fano factor, which breaks down the interpretation of normalized noise. In this case, a more precise definition would be to consider the magnitude of the average current instead.}. 
For a pure Poisson process, the Fano factor is always equal to one. For a deterministic process, it vanishes. 
Using the decompositions in Eq.~\eqref{J-exc} and~\eqref{D-exc} we can express the Fano factor solely in terms of excursion-related quantities: 
\begin{equation}\label{fano_components}
\begin{aligned}
    F_Q &= \frac{{\rm var}(\hat{Q})}{E(\hat{Q})} + \frac{\Delta^2}{\mu^2} E(\hat{Q}) - \dfrac{2}{\mu}{\rm cov}(\hat{Q},\hat{T})
    \\[0.2cm]
    &= F_1 + F_2 + F_3.
\end{aligned}    
\end{equation}
This provides three different mechanisms that contribute to the fluctuation of the steady-state counting observable. 
The first term is the Fano factor of the excursion counting observable only; i.e., it contains only the fluctuations associated to $\hat{Q}$ during an excursion. 
The second term is related to the noise-to-signal ratio $\Delta^2/\mu^2$ of the cycle times [cf. Eqs.~\eqref{mu} and~\eqref{delta2}]. 
It therefore describes how fluctuations in the cycle duration affect the steady-state current. 
Finally, the third term refers to the statistical correlations between the excursion times and the counting observable (notice that because $\hat{\tau}$ and $\hat{T}$ are statistically independent, ${\rm cov}(\hat{Q},\hat{T}) \equiv {\rm cov}(\hat{Q},\hat{T}^{\rm cyc})$. 

\begin{figure}  
    \centering
    \includegraphics[width=1.0\linewidth]{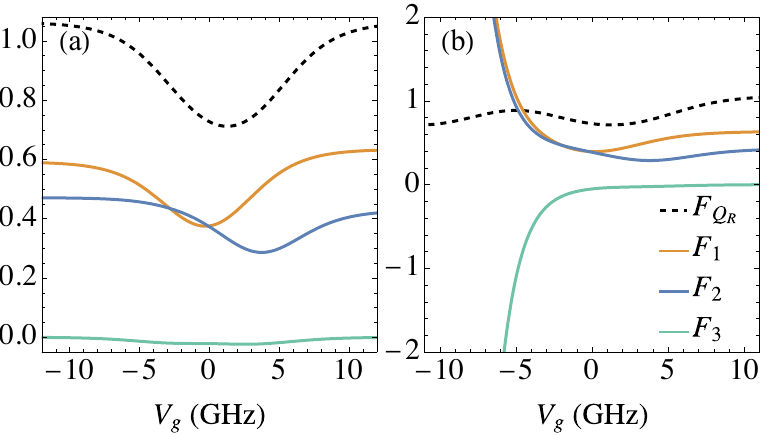}
    \caption{Fano factor of the particle current as a function of $V_g$ (a) with Coulomb blockade and (b) without Coulomb blockade for $U = 10$. Each curve represents a different contribution towards the diffusion coefficient.
    Other parameters (in units of GHz): $V_{sd}=-7$, $\geff = 3.2$, $\gamma = 0.63$, and $k_BT = 2$.}
    \label{fig:fano}
\end{figure}

In Fig.~\ref{fig:fano} we plot each component of Eq.~\eqref{fano_components} for the case of the particle current observable $\hat{Q}_R$, as a function of $V_g$ for fixed $V_{sd}$.  
Fig.~\ref{fig:fano}(a) considers the blockade regime, whereas Fig.~\ref{fig:fano}(b) does not.
In the blockade regime [Fig.~\ref{fig:fano}(a)] all contributions play a significant role.
Conversely, in the non-blockaded regime the Fano factor itself is similar, but the individual contributions are wildly different, and they become very large for negative $V_g$. 
The reason behind this is in the structure of the excursions in the presence or absence of the blockaded regime.
If we assume the system is blockaded, an excursion can have only three possible outcomes for the particle current (as characterized in Section.~\ref{subsec:suc-fail-dis}). 
But when double occupancy is allowed, the excursions may have sub-cycles which add a net positive or negative charge. 
Consequently, the support of $\hat{Q}_R$ becomes the set of all natural numbers.
This introduces a strong correlation between excursion times and $\hat{Q}_R$; namely, long excursions tend to correspond to larger values of $|\hat{Q}_R|$. 
This is clear in Fig.~\ref{fig:fano}(b), where we see that $F_3$ becomes negative and large in magnitude. Since $F_3 \propto -{\rm cov}(\hat{Q},\hat{T})$, this translates into $\hat{Q}_R$ and $\hat{T}$ being positively correlated significantly. 

\subsection{Thermo-kinetic uncertainty relations}\label{ssec:TKUR}

Recently, various thermodynamic bounds have been derived which establish a trade-off between precision and dissipation, as measured by the entropy production~\cite{Gingrich2016, Hasegawa2019a, Hasegawa2019b, Horowitz2019, Vo2022, Di_Terlizzi2018, Campbell2025,Prech2025, Timpanaro2019}.  
The first one discovered was the \emph{thermodynamic uncertainty relation} (TUR)~\cite{Barato2015, Hasegawa2019b, Hasegawa2019a, Horowitz2019}, that can be expressed as
\begin{equation}\label{eq:tur}
    \frac{D_Q}{J_Q^2} \geq \dfrac{2}{J_\Sigma},
\end{equation}
a result which holds for any counting observable $\hat{Q}$ with $\nu_{xy} = -\nu_{yx}$.
The second inequality we consider is the \emph{kinetic uncertainty relation} (KUR)~\cite{Vo2022, Di_Terlizzi2018}, which is given by
\begin{equation}\label{eq:kur}
    \frac{D_Q}{J_Q^2} \geq \dfrac{1}{J_\mathcal{A}}
\end{equation}
where $J_\mathcal{A}$ is the steady-state dynamical activity [Eq.~\eqref{eq:JA}].
The KUR also holds for counting observables that are not anti-symmetric.
Finally, we also discuss the \emph{clock uncertainty relation} (CUR),  recently developed by some of us~\cite{Prech2025}, and which provides a tighter bound to the KUR
\begin{equation}
\label{eq:cur}
      \frac{D_Q}{J_Q^2} \geq \mathcal{T} \geq \frac{1}{J_\mathcal{A}},
\end{equation}
where 
\begin{equation}
    \mathcal{T} = \sum_x \frac{p_x^{\rm ss}}{\Gamma_x},
\end{equation}
is called the average residual time (compare $\mathcal{T}$ with the definition of the dynamical activity $J_\mathcal{A}$ in Eq.~\eqref{eq:JA}).

\begin{figure}
    \centering
    \includegraphics[width=1.0\linewidth]{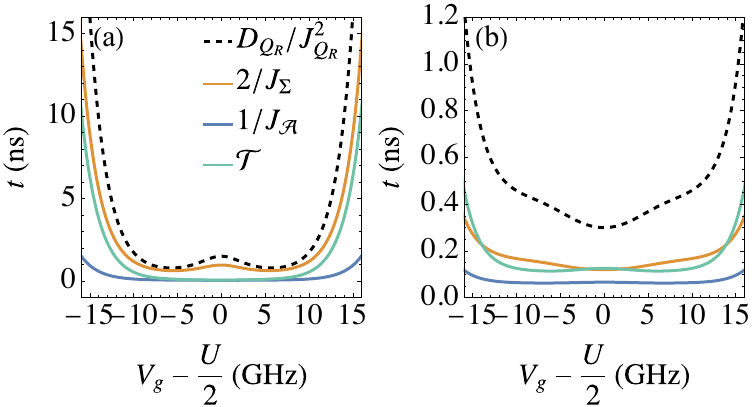}
    \caption{Current precision of the transport observable and the three uncertainty relations discussed here as a function of $V_g$, for (a) $V_{sd} = 7$ and (b) $V_{sd} = -20$. 
    Here we consider fixed values (in units of GHz) of $\geff = 3.2$, $\gamma = 0.63$, $k_BT = 2$, and $U = 10$.}
    \label{fig:uncertainty-relations}
\end{figure}

In Fig.~\ref{fig:uncertainty-relations} we plot the three uncertainty relations above as a function of $V_g$, for two different values of source-drain bias $V_{sd}$. 
For the left-hand side, we use the charge transport observable.
The results, however, would be identically the same for the entropy production (and any other charge-like observable) due to the results in Appendix~\ref{app:conservation}. 
The CUR is always tighter than the KUR. However, CUR and TUR tend to compete, depending on the parameter regime.
For small  bias $|V_{sd}|$  (Fig.~\ref{fig:uncertainty-relations}(a)) the TUR is the tightest uncertainty relation throughout the entire range of gate voltages.
Conversely, for larger $|V_{sd}|$ the TUR and CUR compete on which is the tightest (Fig.~\ref{fig:uncertainty-relations}(b)).
This behavior can be understood by looking at the corresponding time scales. 
The region of very negative $V_g$ corresponds to the long excursion durations, whereas the region of very positive $V_g$ corresponds to the long residence times. 
In both regions, as illustrated by Fig.~\ref{fig:J-entropy-A}(b), there is almost no entropy production, so $2/J_\Sigma$ diverges and dissipation becomes the binding constraint on precision.
The same reasoning applies inside the Coulomb diamond, where the entropy production nearly vanishes while the activity is maximal; only in a small region around $V_g = U/2$ does the CUR become tighter than the TUR.

\section{Conclusion}
\label{sec:conclusion}

The stochastic excursion framework developed in Refs.~\cite{Fiusa2025,Fiusa2025a} provides new insights on the stochastic dynamics of non-equilibrium systems, as it allows one to decompose long stochastic trajectories into physically-motivated sub-trajectories (the excursions). 
This allows one to gain new insights into how fluctuations affect relevant observables, such as particle currents or entropy production. 
Our goal in this paper was to apply this formalism to the problem of a double quantum dot connected to two leads. 
These devices are now routinely studied as quantum thermal machines, due to their thermoelectric properties. 
However, their small sizes make them amenable to significant fluctuations. 
A proper understanding of the thermodynamics of DQDs therefore requires going beyond just average currents, and also understanding fluctuations. 
The excursion framework allowed us to do that, and to reveal new features that govern the DQDs' non-equilibrium dynamics. 
In particular, there are two main insights that follow from this work.

First, analyzing the distribution of counting observable outcomes per excursion provides information about the distribution of failures (see Figs.~\ref{fig:zones-psuc-pfail-pdis} and~\ref{fig:psuc-pfail-pdis}); i.e., events that do not have any thermodynamic outcome. 
Moreover, the probability distribution of outcomes gives rise to a counterintuitive behavior, where depending on the parameter range, it may be beneficial to increase the number of disasters at the cost of significantly suppressing fails, amounting to an overall larger net transport current. Such trade-offs are central to transport and thermodynamics in stochastic systems. While experiments ultimately reveal the answers, a robust theoretical framework is needed to assess these probabilities, and this is one of the features accomplished by the stochastic excursion framework.

The second main insight is the decomposition of noise in terms of specific contributions associated to the excursion [Eq.~\eqref{D-exc}]. 
We analyzed this in terms of the Fano factor [Eq.~\eqref{eq:fano}] in Fig.~\ref{fig:fano}, which uncovered a non-trivial interplay between different noise sources for different parameter regimes. 
In particular, the comparison of the Fano factor decomposition with and without the Coulomb blockade assumption suggests a striking behavior for fluctuations in double quantum dots. 
Although the overall noise contribution is similar, the decomposition is vastly different, and analyzing those different sources may provide hints at the design of useful transport protocols and how the repulsion between excitations in the dots affects transport.

This paper provided a detailed case study of the stochastic excursions framework in an experimentally-motivated system. 
There are various follow-up questions.
In particular, having more states would allow for different decompositions for regions $A$ and $B$, possibly allowing $A$ to have more than one state.
This would unlock the study of correlations between excursions, which can be useful when taking into account memory effects.
In particular, the study of memory-based feedback has been quite active~\cite{rosal2025deterministicequationsfeedbackcontrol, rosal2025deterministicequationsfeedbackcontrol2,rosal2025deterministicequationsfeedbackcontrol3,prech2025quantumthermodynamicscontinuousfeedback, Annby_Andersson2022, Prech2024a} and it would be interesting to explore what kinds of insights a sub-trajectory based framework could bring to this context.
Another promising direction would be to investigate how the decomposition of counting observables at individual excursions can sharpen up bounds on precision or dissipation, in order to refine uncertainty relations.
This investigation is intimately related to our understanding of nonequilibrium processes.
And lastly, it would be of interest to explore the consequences of adding coherent transport and understanding the effect of coherence within such trajectory framework. 

\begin{acknowledgments}
GF acknowledges fruitful discussions with Jiheng Duan and Chaitanya Murthy.
This research is primarily supported by the U.S. Department of Energy (DOE), Office of Science, Basic Energy Sciences (BES) under Award No. DE-SC0025516. 
PH is supported by the European Union (ERC-SuperStoc-101117322).
\end{acknowledgments}

\appendix 

\section{Analytical results under the Coulomb blockade regime}
\label{app:analytical}
 
The excursion toolkit developed in Refs.~\cite{Fiusa2025, Fiusa2025a} provides closed-form, analytical formulas that can be readily applied to several results we discussed in the main text. Throughout the main text, we considered the system outside the blockade region, with all four states. Although analytical formulas can also be derived for this case, they are cumbersome and provide little intuition. We therefore opt to discuss the analytical
results in this Appendix but under the assumption that we are in the blockade regime. For simplicity, we write the results in terms of $\geff$, and we emphasize that the effective coupling is related to the microscopic tunneling amplitude $g$ via Eq.~\eqref{eq:g_eff}. It is convenient to introduce the shorthand
\begin{equation}
  \Lambda := \gamma(1-f_L)(1-f_R),
  \qquad
  \Theta := \Lambda + \geff\,(2-f_L-f_R),
  \label{eq:shorthand}
\end{equation}
in terms of which all the expressions below take a compact form.
 
\paragraph*{Excursion and residence times.---}The average excursion duration straightforwardly follows from Eq.~(24) of Ref.~\cite{Fiusa2025a}:
\begin{equation}
 E(\hat{T})
  = \frac{(2\geff+\gamma)(f_L+f_R) - 2\gamma f_L f_R}
         {\gamma\,(f_L+f_R)\,\Theta},
  \label{eq:ET}
\end{equation}
and the average residence time follows trivially,
\begin{equation}
  E(\hat{\tau}) = \frac{1}{\Gamma_x} = \frac{1}{\gamma(f_L+f_R)}.
  \label{eq:Etau}
\end{equation}
Hence it directly follows that
\begin{equation}
\begin{aligned}
  \mu &= \frac{(2\geff+\gamma)(f_L+f_R) - 2\gamma f_L f_R + \Theta}
             {\gamma\,(f_L+f_R)\,\Theta}\\
      &= \frac{\geff(2+f_L+f_R) + \gamma(1-f_L f_R)}
             {\gamma\,(f_L+f_R)\,\Theta}.
  \label{eq:mu}
\end{aligned}
\end{equation}
The variance of the excursion durations and cycle durations are also straightforward to calculate, but the expressions are rather cumbersome and have been omitted.
 
\paragraph*{Average transport.---}The average value of the transport observable follows directly from
Eq.~(32) of Ref.~\cite{Fiusa2025a}. It reads
\begin{equation}
  E(\hat{Q}_R)
  = \frac{\geff\,(f_L - f_R)}{(f_L+f_R)\,\Theta}.
  \label{eq:EQR}
\end{equation}
Together with Eq.~\eqref{eq:mu}, the particle current [see Eq.~\eqref{J-exc}] is straightforwardly evaluated. Notice that Eq.~\eqref{eq:EQR} coincides with $P_{\rm suc}-P_{\rm dis}$ obtained from Eqs.~\eqref{eq:psuc} and~\eqref{eq:pdis}, as it must.
 
\paragraph*{Average dynamical activity.---}Similarly, the dynamical activity follows directly from Eq.~(32) of
Ref.~\cite{Fiusa2025a}. It reads
\begin{equation}
\begin{aligned}
  E(\hat{\mathcal{A}})
  &= \frac{2\geff^{2}(f_L+f_R) + 2\gamma^{2}(1-f_L)(1-f_R)(f_L+f_R)
          }
         {\gamma\,(f_L+f_R)\,\Theta}\\
         &+\dfrac{\geff\gamma\big[5(f_L+f_R) - 6 f_L f_R - 2(f_L^{2}+f_R^{2})\big]}{\gamma\,(f_L+f_R)\,\Theta}
  \label{eq:EA}
  \end{aligned}
\end{equation}
and the dynamical activity per unit time is evaluated by combining Eq.~\eqref{eq:JA} with the explicit formula for $\mu$ [Eq.~\eqref{eq:mu}].
 
\paragraph*{Average entropy production.---}The entropy production requires no separate calculation. By the pathwise
identity established in Appendix~\ref{app:proportionality} [Eq.~\eqref{eq:pathwise}], the entropy production within an excursion is a fixed multiple of the transport observable, so that
\begin{equation}
\begin{aligned}
  E(\hat{\Sigma})
  &= \log\!\left[\frac{f_L(1-f_R)}{f_R(1-f_L)}\right]\,
    E(\hat{Q}_R)\\
  &= \log\!\left[\frac{f_L(1-f_R)}{f_R(1-f_L)}\right]
    \frac{\geff\,(f_L-f_R)}{(f_L+f_R)\,\Theta}.
  \label{eq:ESigma}
  \end{aligned}
\end{equation}
Note that $E(\hat{\Sigma})$ vanishes at $f_L = f_R$, as it must at equilibrium, and is nonnegative for any $f_L \neq f_R$ since the logarithm and $(f_L-f_R)$ always carry the same sign. The connection with the entropy production rate is established by dividing the above result by Eq.~\eqref{eq:mu}.
 
\paragraph*{Populations.---}The populations of each dot can be expressed with simple analytical formulas. The population in the left dot reads
\begin{equation}
  p_L = \frac{\geff (f_L+f_R) + \gamma f_L(1-f_R)}
             {\geff(2+f_L+f_R) + \gamma(1-f_L f_R)},
  \label{eq:pL}
\end{equation}
and the population in the right dot is obtained by exchanging
$L \leftrightarrow R$,
\begin{equation}
  p_R = \frac{\geff (f_L+f_R) + \gamma f_R(1-f_L)}
             {\geff(2+f_L+f_R) + \gamma(1-f_L f_R)}.
  \label{eq:pR}
\end{equation}
From normalization, the probability of having both dots empty is
\begin{equation}
  p_{00} = 1 - p_L - p_R
         = \frac{\Theta}{\geff(2+f_L+f_R) + \gamma(1-f_L f_R)},
  \label{eq:p00}
\end{equation}
which, as a consistency check, is precisely $E(\hat{\tau})/\mu$, that is, the fraction of time the system spends
residing in region $A$. Although the Coulomb blockade regime is a significant simplification, it is notable that the populations of each dot are expressed with simple closed-form formulas.

\section{Conservation laws at the level of individual excursions}
\label{app:conservation}
Here we show that the excursion framework makes the statement of proportionality of currents sharp: within a single excursion, the two transport observables satisfy $\hat{Q}_L = -\hat{Q}_R$ as random
variables, and not merely on average. We then use the same mechanism to show that the entropy production is a fixed multiple of the charge current on every excursion. Both results hold for the full four-state model and for
arbitrary Coulomb repulsion $U$ (the blockade limit is not required).
 
\subsection{Charge conservation within an excursion}
\label{app:charge}
The transport observables to the left and right baths are defined by the
weight matrices
\begin{equation}
  \nu_{Q_L} =
  \begin{pmatrix}
    0 & 1 & 0 & 0\\
   -1 & 0 & 0 & 0\\
    0 & 0 & 0 & 1\\
    0 & 0 &-1 & 0
  \end{pmatrix},
  \qquad
  \nu_{Q_R} =
  \begin{pmatrix}
    0 & 0 & 1 & 0\\
    0 & 0 & 0 & 1\\
   -1 & 0 & 0 & 0\\
    0 &-1 & 0 & 0
  \end{pmatrix},
  \label{eq:nuLnuR}
\end{equation}
in the ordering $x = (00,10,01,11)$ used throughout, with $\nu_{Q_R}$ defined in Eq.~\eqref{nu_current} in the main text. Consider now the counting observable $\hat{Q}_n$ that registers a particle leaving the system, irrespective of which bath it goes to. Its weight matrix is simply the sum
\begin{equation}
  \nu_n = \nu_{Q_L} + \nu_{Q_R} =
  \begin{pmatrix}
    0 & 1 & 1 & 0\\
   -1 & 0 & 0 & 1\\
   -1 & 0 & 0 & 1\\
    0 &-1 &-1 & 0
  \end{pmatrix}.
  \label{eq:nun}
\end{equation}
Region $A$ has only one state. Therefore, within a single excursion, the total value of $\hat{Q}_n$ should always be identically zero: whatever particles enter the system midway through the excursion must necessarily
leave it before the excursion ends. The corresponding joint excursion distribution [Eq.~\eqref{Pqt}] must then have the degenerate form
\begin{equation}
  P_{Q_n, T}(q_n,t) = P_T(t)\,\delta(q_n)
  \label{eq:degenerate}
\end{equation}
i.e. $q_n$ always takes the same value. In other words, at the level of the random variable, $\hat{Q}_n = 0$. Since $\hat{Q}_n = \hat{Q}_L + \hat{Q}_R$ by construction, this is equivalent to
\begin{equation}
  \hat{Q}_L = -\hat{Q}_R
  \label{eq:QLQR}
\end{equation}
on every individual excursion. This is a \emph{strong} conservation law, in the sense that it constrains individual realizations and not only averages.
 
It is worth emphasizing why the excursion structure is essential here. If one instead looks at counting observables along a long trajectory, it is not necessarily true that $\hat{Q}_n = 0$, because if we start from a state with a definite number of particles in the system, after an arbitrary time $t$ the
occupation may well be different. What survives for long-time statistics is the weaker statement that $\hat{Q}_n(t)$ does not grow with $t$.

It is also possible to prove Eq.~\eqref{eq:QLQR} from the algebraic properties of the tilted matrix. And this highlights some useful features concerning conserved quantities at the excursion level. 
We start from the joint distribution of a counting observable and the excursion time, Eq.~\eqref{Pqt} of the main text,
\begin{equation}
  P(q,t) = \int_{-\infty}^{\infty}\frac{d\xi}{2\pi}\,
  \frac{\bra{x_A} W_{AB\xi}\, e^{\mathbb{W}_{B\xi}t}\, W_{BA\xi} \ket{x_A}\,
        e^{iq\xi}}
       {\bra{x_A} W_{AB}(-\mathbb{W}_B^{-1}) W_{BA} \ket{x_A}},
  \label{eq:Pqt}
\end{equation}
applied to $\hat{Q}_n$. 
Our goal is to prove that the only $\xi$ dependence in the integrand should be the factor of $e^{i q \xi}$. 
If this is the case, the integral will produce a delta function and Eq.~\eqref{eq:degenerate} follows. Using the tilting prescription of Eq.~\eqref{tilted_transition_matrix}, $(\mathbb{W}_\xi)_{xy} = \mathbb{W}_{xy}e^{-i\nu_{xy}\xi}$ for $x\neq y$,
the weights~\eqref{eq:nun} give
\begin{equation}
  \mathbb{W}_\xi =
  \begin{pmatrix}
    0 & \gamma(1-f_L)e^{-i\xi} & \gamma(1-f_R)e^{-i\xi} & 0\\[2pt]
    \gamma f_L e^{i\xi} & 0 & \geff & \gamma(1-\tilde{f}_R)e^{-i\xi}\\[2pt]
    \gamma f_R e^{i\xi} & \geff & 0 & \gamma(1-\tilde{f}_L)e^{-i\xi}\\[2pt]
    0 & \gamma\tilde{f}_R e^{i\xi} & \gamma\tilde{f}_L e^{i\xi} & 0
  \end{pmatrix} - \Gamma .
  \label{eq:Wxi}
\end{equation}
The diagonal is untouched because $\nu_{xx}=0$.
The key observation is that the phases in Eq.~\eqref{eq:Wxi} are not independent of one another: each entry carries a phase determined solely by the two states it connects. Indeed, let $N_x$ denote the number of particles
in state $x$, i.e.
\begin{equation}
  N_{00} = 0, \qquad N_{10} = N_{01} = 1, \qquad N_{11} = 2 .
  \label{eq:Nx}
\end{equation}
Looking at Eq.~\eqref{eq:nun} entry by entry, one verifies that
\begin{equation}
  (\nu_n)_{xy} = N_y - N_x
  \qquad\text{for every pair } (x,y) \text{ with } W_{xy} \neq 0 .
  \label{eq:gradient}
\end{equation}
A term like 
$(\nu_n)_{11,00}$ does not satisfy Eq.~\eqref{eq:gradient}, but this does not matter since the transitions between state $00$ and $11$ do not take place.
Alternatively, we could also define a new counting matrix 
\begin{equation}
    \nu_n' = \begin{pmatrix}
    0 & 1 & 1 & 2\\
   -1 & 0 & 0 & 1\\
   -1 & 0 & 0 & 1\\
    -2 &-1 &-1 & 0
  \end{pmatrix} = N_y - N_x,
\end{equation}
This is now exactly $N_y - N_x$; and the factors of $\pm 2$ will not really matter since the corresponding value of $\mathbb{W}_{xy}$ are zero anyway. 

Eq.~\eqref{eq:gradient} means that 
\begin{equation}
  (\mathbb{W}_\xi)_{xy}
  = \mathbb{W}_{xy}\, e^{-i(N_y - N_x)\xi}
  = e^{iN_x\xi}\; \mathbb{W}_{xy} \; e^{-iN_y\xi}.
  \label{eq:entrywise}
\end{equation}
And if we define the diagonal matrix 
\begin{equation}
  G_\xi := \operatorname{Diag}\!\left(e^{iN_x\xi}\right)
         = \operatorname{Diag}\!\left(1,\; e^{i\xi},\; e^{i\xi},\; e^{2i\xi}\right),
  \label{eq:Gxi}
\end{equation}
then 
\begin{equation}
 \mathbb{W}_\xi = G_\xi\, \mathbb{W}\, G_\xi^{-1},
  \label{eq:similarity}
\end{equation}
so, for the tilting weights~\eqref{eq:nun},  $\mathbb{W}_\xi$ and $\mathbb{W}$, are related by a similarity transformation.
 
Since $G_\xi$ is diagonal, it inherits the block structure of Eq.~\eqref{block_W}; namely $G_\xi = \operatorname{Diag}(G_A, G_B)$, with
\begin{equation}
  G_A = 1, \qquad
  G_B = \operatorname{Diag}\!\left(e^{i\xi}, e^{i\xi}, e^{2i\xi}\right).
  \label{eq:GAGB}
\end{equation}
Hence 
\begin{equation}
\begin{aligned}
  &W_{AB\xi} = G_A\, W_{AB}\, G_B^{-1},\\
  \qquad
 &W_{BA\xi} = G_B\, W_{BA}\, G_A^{-1},\\
  \qquad
  &\mathbb{W}_{B\xi} = G_B\, \mathbb{W}_B\, G_B^{-1}.
  \label{eq:blocks}
  \end{aligned}
\end{equation}
Since similarity transformations satisfy  $e^{G Y G^{-1}} = G e^{Y} G^{-1}$, we have that 
\begin{equation}
  e^{\mathbb{W}_{B\xi}t} = G_B\, e^{\mathbb{W}_{B}t}\, G_B^{-1}.
  \label{eq:expblock}
\end{equation}
Combining the three factors in the numerator of Eq.~\eqref{eq:Pqt}, the $G_B$ factors cancel pairwise,
\begin{align}
  W_{AB\xi}\, e^{\mathbb{W}_{B\xi}t}\, W_{BA\xi}
  &= \big(G_A W_{AB} G_B^{-1}\big)
     \big(G_B e^{\mathbb{W}_B t} G_B^{-1}\big)
     \big(G_B W_{BA} G_A^{-1}\big)
  \nonumber\\[2pt]
  &= G_A \Big[ W_{AB}\, e^{\mathbb{W}_B t}\, W_{BA} \Big] G_A^{-1}
   = W_{AB}\, e^{\mathbb{W}_B t}\, W_{BA},
  \label{eq:cancellation}
\end{align}
where in the last step we used that region $A$ contains a single state, so that $G_A$ is a scalar and cancels against its own inverse. The numerator is therefore independent of $\xi$, so the integral in Eq.~\eqref{eq:Pqt}
gives $\delta(q_n)$, and $\hat{Q}_n = 0$ at the level of the random variable. This proves Eq.~\eqref{eq:QLQR}.

This result provides a mathematically simple way of establishing conservation laws at the level of individual excursions. 
Namely, any weight matrix $\nu_{xy}$
that can be written as 
\begin{equation}
    \nu_{xy} = N_y - N_x,
    \label{eq:nu_Nx_Ny}
\end{equation}
produces a  counting observable that is conserved at the level of individual excursions. 
 
\subsection{A second conserved quantity}
\label{app:Q11}

There exists one further observable in this model, which is also conserved, and which can be used to show that all thermodynamic currents are proportional to each other.
Consider the observable $\hat{Q}_{11}$ that counts the net number of times the system leaves the doubly-occupied state. That is, it adds $+1$ for every transition out of $\ket{11}$ and $-1$ for every transition into it. Its
weights are given by
\begin{equation}
  \nu_{11} =
  \begin{pmatrix}
    0 & 0 & 0 & 0\\
    0 & 0 & 0 & 1\\
    0 & 0 & 0 & 1\\
    0 &-1 &-1 & 0
  \end{pmatrix}.
  \label{eq:nu11}
\end{equation}
This can also be written in the form~\eqref{eq:nu_Nx_Ny}, but with 
\begin{equation}
    N_{00} = N_{01} = N_{10} = 1,\quad N_{11} =1.
\end{equation}
Hence the same proof of the previous section carries over and we conclude that 
\begin{equation}
  \hat{Q}_{11} = 0
  \label{eq:Q11zero}
\end{equation}
on every individual excursion.
This conservation law can also be understood intuitively.
An excursion begins and ends in $\ket{00}$, so the system cannot be sitting in $\ket{11}$ at either endpoint, and it must therefore leave the doubly-occupied state precisely as
many times as it enters it.

\subsection{Proportionality between entropy production and charge current}
\label{app:proportionality}
 
We can now show that the entropy production carries no information beyond the charge current. Let us introduce the notation
\begin{equation}
  \zeta_a := \log\!\left(\frac{1-f_a}{f_a}\right),
  \qquad
  \tilde{\zeta}_a := \log\!\left(\frac{1-\tilde{f}_a}{\tilde{f}_a}\right),
  \qquad a = L,R,
  \label{eq:zetas}
\end{equation}
so that the weights of entropy production [Eq.~\eqref{nu_entropy}] read
\begin{equation}
  \nu_\Sigma =
  \begin{pmatrix}
    0 & \zeta_L & \zeta_R & 0\\
    -\zeta_L & 0 & 0 & \tilde{\zeta}_R\\
    -\zeta_R & 0 & 0 & \tilde{\zeta}_L\\
    0 & -\tilde{\zeta}_R & -\tilde{\zeta}_L & 0
  \end{pmatrix}.
  \label{eq:nuSigmaZ}
\end{equation}
Using the definitions of $f_a$ and $\tilde{f}_a$ in Sec.~\ref{sec:formalism}, which differ only by the replacement
$V_{g_a} \to V_{g_a} + U$, one finds directly that
\begin{equation}
  \tilde{\zeta}_a = \zeta_a + \beta U .
  \label{eq:zetatilde}
\end{equation}
This is the only property we need. Comparing Eq.~\eqref{eq:nuSigmaZ} with the weight matrices $\nu_{Q_L}$, $\nu_{Q_R}$ and $\nu_{11}$ of Eqs.~\eqref{eq:nuLnuR} and~\eqref{eq:nu11}, entry by entry, we may write
\begin{equation}
  \nu_\Sigma = \zeta_L\, \nu_{Q_L} + \zeta_R\, \nu_{Q_R}
             + \beta U\, \nu_{11} .
  \label{eq:nuSigmaDecomp}
\end{equation}

Note that Eq.~\eqref{eq:nuSigmaDecomp} is a relation between weight matrices, and it translates immediately into a relation between the corresponding counting observables within a single excursion,
\begin{equation}
  \hat{\Sigma} = \zeta_L \hat{Q}_L + \zeta_R \hat{Q}_R
               + \beta U\, \hat{Q}_{11} .
  \label{eq:SigmaDecomp}
\end{equation}
The last term vanishes identically by Eq.~\eqref{eq:Q11zero}, and the first two can be factored into one by using the charge conservation law $\hat{Q}_L = -\hat{Q}_R$ of Eq.~\eqref{eq:QLQR}. We then obtain
\begin{equation}
\hat{\Sigma} = \big(\zeta_R - \zeta_L\big)\,\hat{Q}_R
   = \log\!\left[\frac{f_L(1-f_R)}{f_R(1-f_L)}\right]\hat{Q}_R
  \label{eq:pathwise}
\end{equation}
as an identity between random variables, which hold at individual excursions, even  for the full four-state model, and for any value of $U$.
 
It is worth noting where the assumptions of the model enter. The whole argument relies on Eq.~\eqref{eq:zetatilde}, i.e. on the fact that the shift $\beta U$ between $\zeta_a$ and $\Tilde{\zeta}_a$ is the same for $a = L, R$. This is a consequence of the two reservoirs sharing a common inverse temperature $\beta_L = \beta_R = \beta$. Had the temperatures differed, the two shifts would be $\beta_L U$ and $\beta_R U$, and then the remaining term would no longer be proportional to $\nu_{11}$, and no relation of the form~\eqref{eq:pathwise} would follow. Conversely, the equality of the gate voltages is not needed anywhere above, all it does is affect the numerical value of $\zeta_R - \zeta_L$, which for $V_{g_L}=V_{g_R}$ and $\mu_{L,R} = \mp V_{sd}/2$ reduces to $-\beta V_{sd}$.
 
\subsection{Consequences}
\label{app:consequences}
 
Since Eq.~\eqref{eq:pathwise} relates the random variables themselves, every statistical property of $\hat{\Sigma}$
follows from the corresponding property of $\hat{Q}_R$. Let us denote
$\zeta := \zeta_R - \zeta_L$ for brevity, then
\begin{equation}
\begin{aligned}
  &E(\hat{\Sigma}) = \zeta\,E(\hat{Q}_R),\\
  \quad
  &\text{var}(\hat{\Sigma}) = \zeta^{2}\,\text{var}(\hat{Q}_R),\\
  \quad
  &\text{cov}(\hat{\Sigma},\hat{T})
    = \zeta\,\text{cov}(\hat{Q}_R,\hat{T}),
  \label{eq:moments}
  \end{aligned}
\end{equation}
and hence, through Eqs.~\eqref{J-exc} and \eqref{D-exc}, we conclude
\begin{equation}
  J_\Sigma = \zeta\, J_{Q_R},
  \qquad
  D_\Sigma = \zeta^{2}\, D_{Q_R} .
  \label{eq:currents}
\end{equation}
Since the entropy production enters the uncertainty relations only through these two quantities, the precision
$D_Q/J_Q^{2}$ appearing on the left-hand side of Eqs.~\eqref{eq:tur} and \eqref{eq:kur} is the same
whether one uses the charge current or the entropy production.

Finally, we note that the argument is not restricted to the entropy production. Any counting observable whose weights decompose as in Eq.~\eqref{eq:nuSigmaDecomp} is likewise a multiple of $\hat{Q}_R$. This is not, however, true of every thermodynamic current. For example, the heat delivered to an individual lead: consider the excursion $\ket{00} \to \ket{10} \to \ket{11} \to \ket{01} \to \ket{00}$, this has $\hat{Q}_R = 0$ and $\hat{\Sigma} = 0$, but it transfers an energy $U$ from the right reservoir to the left one. 
Therefore, the double quantum dot model supports two independent currents per excursion, charge and heat, and it is the condition $\beta_L = \beta_R = \beta$ that makes the entropy production behave like a charge.
Under the blockade, the state $\ket{11}$ is no longer accessible, hence transitions involving this state become forbidden and then all thermodynamic currents indeed become proportional.


\bibliography{letter}

\end{document}